\documentclass[]{raa}            

\usepackage{graphicx,times}             
\usepackage{multirow}

\begin{document}

   \title{Spectral classification of stars based on LAMOST spectra
}

   \volnopage{Vol.0 (200x) No.0, 000--000}      
   \setcounter{page}{1}          

   \author{Chao Liu
      \inst{1}
   Wen-Yuan Cui\inst{2}
   Bo Zhang\inst{1}
   Jun-Chen Wan\inst{1}
    Li-Cai Deng\inst{1}
    Yonghui Hou\inst{3}
    Yuefei Wang\inst{3}
    Ming Yang\inst{1}
    \and Yong Zhang\inst{3}
   }

   \institute{Key laboratory of Optical Astronomy, National Astronomical Observatories, Chinese Academy of Sciences,
             Beijing 100012, China; {\it liuchao@bao.ac.cn}\\
               \and
               Department of Physics, Hebei Normal University, Shijiazhuang 050024, China\\
               \and Nanjing Institute of Astronomical Optics \& Technology, National Astronomical Observatories, Chinese Academy of Sciences, Nanjing 210042, China\\
 }

   \date{Received~~2009 month day; accepted~~2009~~month day}

\abstract{In this work, we select the high signal-to-noise ratio spectra of stars from the LAMOST data and map their MK classes to the spectral features. The equivalent widths of the prominent spectral lines, playing the similar role as the multi-color photometry, form a clean stellar locus well ordered by MK classes. The advantage of the stellar locus in line indices is that it gives a natural and continuous classification of stars consistent with either the broadly used MK classes or the stellar astrophysical parameters. We also employ a SVM-based classification algorithm to assign MK classes to the LAMOST stellar spectra. We find that the completenesses of the classification are up to 90\% for A and G type stars, while it is down to about 50\% for OB and K type stars. About 40\% of the OB and K type stars are mis-classified as A and G type stars, respectively.  This is likely owe to the difference of the spectral features between the late B type and early A type stars or between the late G and early K type stars are very weak. The relative poor performance of the automatic MK classification with SVM suggests that the directly use of the line indices to classify stars is likely a more preferable choice.
\keywords{techniques: spectroscopic --- stars: general---stars: fundamental parameters---stars: statistics---Galaxy: stellar contents }
}

   \authorrunning{Liu et al.}            
   \titlerunning{Spectral classification of stars}  

   \maketitle

%
%
\section{Introduction} \label{sect:intro}
The classification of the normal stars plays important roles not only in the understanding of the stellar physics, but also in the study of the overall structure and evolution of the Milky Way. MK classification (Morgan \& Keenan~\cite{MK73}) is one of the most broadly used systems based on the spectral features of a small number of standard stars. Compared to the mapping of the stars directly using the effective temperature, surface gravity, and chemical abundances, the MK classification is simple and effective. A usual procedure to process a spectrum in a spectroscopic survey is to firstly assign the MK classes to the spectra and then to estimate the stellar astrophysical parameters using the MK classes as the start points (e.g. Luo et al.~\cite{luo15}). Spectral classifications are also very helpful in the targeting for the follow-up studies. For instance, in order to select the blue horizontal branch stars from the whole dataset, one might firstly select all A type stars to reduce the size of the sample; in order to study the circumstellar environment of the young massive stars, one needs to firstly select the OB type stars from the full sample; or in order to search for the AGB stars, one has to firstly select the M giant stars from the database.  

Alternatively, the stars can be classified based on the color indices. Now-a-days, billions of stars have accurate multi-band photometry covering from UV to infrared bands, e.g. GALEX (Bianchi et al.~\cite{galex}), SDSS (Ahn et al.~\cite{ahn14}), PanSTARRS (Tonry et al.~\cite{panstarrs}), 2MASS (Skrutskie et al.~\cite{2mass}), WISE (Wright et al.~\cite{wise}) etc. which provide abundant information of the stellar astrophysical parameters. For instance, Covey et al.~(\cite{covey07}) mapped the stars with different types in SDSS+2MASS multi-color space and showed a clear continuous stellar locus, on which any reasonable stellar classifications can be set up, including the well known MK class system. The biggest advantage of the continuous stellar locus in color space is that it naturally reflects how the spectral energy distribution varies with the stellar astrophysical parameters, such as the effective temperature, the surface gravity, and the metallicity. Therefore, it is very important in either the researches on the stellar evolution or the overall features of the Milky Way. In fact, in the context of the survey with millions to billions of stars, the color-based stellar locus may be more effective and straightforward in the stellar classifications (see their applications in Yanny et al.~\cite{yanny00}, Majewski et al.~\cite{majewski03}, Yanny et al.~\cite{yanny09} etc.)  However, when one goes into deep sky, especially along the Galactic mid-plane, most of the photometric color indices of stars are reddened by the absorption and scattering of the interstellar medium. Although some remarkable works has been done (Schlegel et al.~\cite{schlegel98}, Schlafly et al.~\cite{schlafly15}, Chen et al.~\cite{chen14}), the knowledge of the 3 dimensional reddening distribution of the Milky Way is still very limited, leading to certain systematics varies with lines of sight and distances in the multi-color index-based stellar classifications. Moreover, in most cases, the color index is an integration of the spectrum over a wide range of wavelength, details showing in the spectral lines are smoothed out. Therefore, in general, color index-based classifications of stars cannot completely take the place of the spectra-based classifications.

Most of the known MK types of the stars are classified by manually comparing the spectra with a small sets of standard stars (e.g. the samples in Corbally, Gray \& Garrison~\cite{corbally94}), which is not efficient when the sample is huge and not always reliable. Although efforts has been made to automize the MK classification by developing an automatic software (e.g., Gray \& Corbally~\cite{gray14}), it is still a non-trivial task since the real stellar spectra are not only sensitive to the effective temperature and luminosity, but also dependent on the elemental abundances. Moreover, in a large spectroscopic survey, the spectra may be in the wide range of signal-to-noise ratio, making the spectral features not always as clear as the small set of well observed high-quality standard spectra. For such a large spectroscopic survey, the mis-classification in the template-matching techniques based on some standard stars may be significant for low signal-to-noise ratio data. And it may subsequently affect the effort of the searching of the peculiar and rare objects.

Other efforts of the star classification based on automatic algorithms have been done in the past twenty years by various works. These algorithms include metric-distance technique (e.g. LaSala~\cite{lasala94}), artificial neural networks (e.g. Bailer-Jones~\cite{bailerjones97}), fuzzy logic methods (e.g. Carricajo et al.~\cite{carricajo04}) etc. It is noted that Bailer-Jones et al.~(\cite{bailerjones08}) and Saglia et al.~(\cite{saglia12}) reported the applications of the support vector machine (SVM) in the star-galaxy-QSO classifications. In general, this new technique can also be used for the classification of stars.

Recently, the LAMOST survey (Cui et al.~\cite{cui12}; Zhao et al.~\cite{zhao12}; Deng et al.~\cite{deng12}) has been collected more than 4 million stellar spectra in its 2nd internal data release (DR2). Unlike SDSS, the LAMOST survey does not have the combined photometry survey with its spectroscopic one, for which the targets are selected from several external photometric catalogs (Carlin et al.~\cite{carlin12}; Yuan et al.~\cite{yuan14}). This makes it difficult to establish the star classification based on the photometric color indices since the multiple input catalogs are not well calibrated. With only the stellar spectra, it is not trivial to automatically classify the stars into different MK types. The LAMOST pipeline (Luo et al.~\cite{luo12}; Luo et al.~\cite{luo15}) runs a cross-correlation based algorithm (correlation function initial; CFI) to assign the MK types to each stellar spectra. However, due to some technical issues (e.g., the noises in the spectra, the interstellar extinction distortion in the continua, and the limit of the synthetic library used in CFI etc.), this classification, which has already appeared in the LAMOST catalog, is not very reliable, especially for O, B, A, and M type stars. Therefore, a robust and reliable automatic classification method suitable for all spectral classes for the LAMOST spectra is anxiously required.

In this work, we map the MK classes to the space of the indices of the prominent spectral lines in the spectra. The line indices naturally form a stellar locus from the hottest to the coolest stars because the smooth transition of these spectral lines with the effective temperature and surface gravity of the stars. In principle, unlike the broadly used discrete MK classes, the line indices can automatically provide a continuous classes, although the elemental abundance may broaden it. Meanwhile, the MK classes or other class systems can be easily mapped to the line index space to find their counterparts. We also employ the SVM to assign the MK classes for the stars and compare it with the line indices-based classification. We suggest that the line indices-based classification is one of the most robust ways to classify the stars in the era of large data.

The paper is organized as below. In section 2, we give a brief introduction of the LAMOST survey and the data selection for the classification. We also give the detailed definition of the indices of more than 20 spectral lines in the rest of this section. In section 3, we show the features of the stellar locus in the space of the line indices and how the locus associates with the MK classes. In section 4, we employ the support vector machine to classify the stars to MK types. We then compare the stellar locus-based with the SVM-based MK classification. We raise discussions in section 5 and draw the short conclusion in section 6.

\section{Data}\label{sect:data}

\subsection{LAMOST and SIMBAD data}\label{sect:datamatch}
The LAMOST telescope, also called Guo Shou Jing telescope, is a 4-m reflected Schmidt telescope with 4000 fibers configured on the 5-degree field of view (Cui et al.~\cite{cui12}; Zhao et al.~\cite{zhao12}). The LAMOST Milky Way survey will finally targets more than 5 million stellar spectra with resolution of R$\sim1800$ in its 5-year observations (Deng et al.~\cite{deng12}; Liu, X-W. et al.~\cite{lxw13}). It seems that it can obtain more spectra than the schedule after the LAMOST team released the DR2 catalog, which contains about 4 million stellar spectra, by the end of 2014. 

We select about 1.52 million stellar spectra with signal-to-noise ratio larger than 20 (which means both the averaged signal-to-noise ratio at $g$ and $i$ band are larger than 20) to investigate how the spectral features vary with the star classes. In order to identify their MK classes, we cross identify them with the SIMBAD catalog (Wenger et al.~\cite{simbad})\footnote{http://simbad.u-strasbg.fr/Simbad} and obtain 3,134 spectra of normal stars with MK classification flags in the SIMBAD catalog. Table~\ref{tab:sample} shows the distribution of the MK classes for the sample. It shows that the sample is distributed in the MK classes in significant imbalance. The stars between late B and early A and the G, K type stars are prominent in the sample. And the main-sequence stars is much more than the giant stars, while the supergiant stars are very rare.

\subsection{Line indices}\label{sect:linedef}

In order to associate the star classes with the spectral features, we measure the line indices of spectral lines instead of using the full spectra. In general, line indices do not request the flux calibration, which is very hard to calibrate in the LAMOST pipeline due to the complicated instrument, e.g., 4000 fibers with different length, 16 spectrographs with slightly different performances etc. They are also very robust against the random noise. Although the sky background is very difficult to be cleanly subtracted from the spectra, the blue part of the spectra are less influenced. Fortunately, most of the well known line indices, e.g. the Lick indices (Worthey et al.~\cite{worthey94}; Worthey \& Ottaviani~\cite{worthey97}) are in blue. 

The principle of the selection of the spectral lines is two-fold. First, the lines should be strong enough that can be effectively detected in the low resolution spectra. Second, the lines should be sensitive to the effective temperature, surface gravity, and metallicity so that they can play roles in the classification. Table~\ref{tab:lineindices} lists all 27 spectral lines used in this work. Most of them are adopted from Lick indices (Worthey et al.~\cite{worthey94}; Worthey \& Ottaviani~\cite{worthey97}; Cohen, Blakeslee \& Ryzhov~\cite{cohen98}). To better separate OB type stars we add three Helium lines. And since CaII K line may also be often used for classification, it is also considered based on the definition by Beers et al.~(\cite{beers99}).

We define the line index in terms of equivalent width ($EW$) with the following equation (Worthey et al.~\cite{worthey94}):
\begin{equation}
EW=\int{(1-{f_{line}(\lambda)\over{f_{cont}(\lambda)}})}{d\lambda},
\end{equation}
where $f_{cont}(\lambda)$ and $f_{line}(\lambda)$ are the fluxes of the continuum and the spectral line, respectively, both of which are functions of the wavelength $\lambda$. The continuum $f_{cont}$ is estimated via linear interpolation of the fluxes located in the pseudo-continuum region on either side of each index bandpass (see table~\ref{tab:lineindices}). The line index under this definition is in $\rm\AA$. {It is noted that the measurement of the equivalent widths of the lines is based on the rest-frame spectra, in which the radial velocities have been corrected. The value of the radial velocity is adopted from the LAMOST catalog. For the spectra with signal-to-noise ratio larger than 20, the median uncertainty of the equivalent widths of the lines are smaller than 0.1$\rm\AA$.}

\begin{figure}
 \begin{center}
 \includegraphics[scale=0.5]{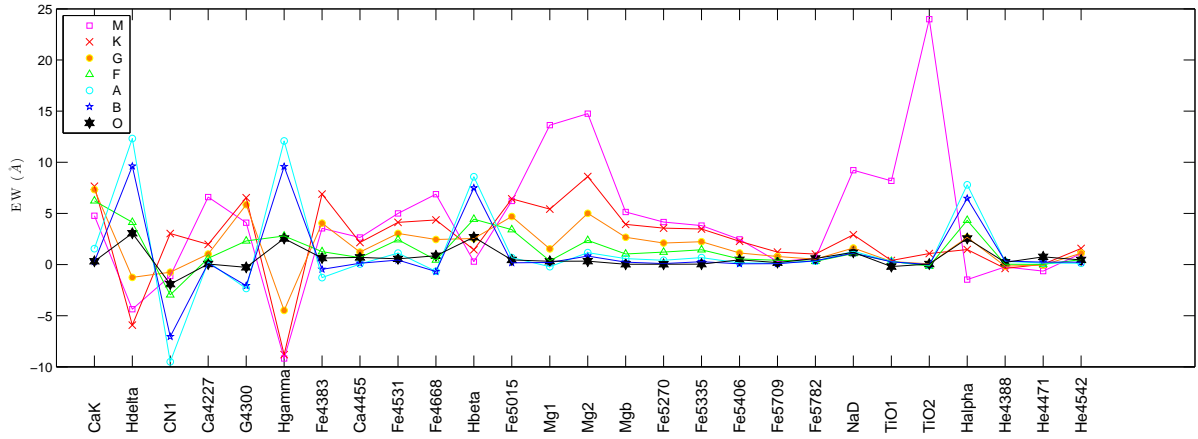}
 \caption{The figure shows the line indices for different MK classes. Each grid in the x-axis corresponds to a spectral line and y-axis indicates the median equivalent width for each line. The colors and symbols codes the O (black crosses), B (blue pentagons), A (cyan large circles), F (green triangles), G (orange small circles), K (red hexagons), and M (Magenta rectangles) types.}\label{fig:lineindices}
 \end{center}
 \end{figure}

Fig~\ref{fig:lineindices} shows the median equivalent width of different spectral lines for each class of stars. It is evident that the spectral lines are not equally sensitive to the stellar classes. All Balmer lines, i.e., H$_\alpha$, H$_\beta$, H$_\gamma$, and H$_\delta$, well separate the classes. The Magnesium lines are also sensitive to the classes, particularly for late type stars. Although the iron lines change not as significant as the Mg lines, they also show clear trend in different classes. Finally, the TiO lines are very sensitive to the M type stars. It seems that lots of the spectral lines are correlated. Hence, we do not need to use all of them for the classification. We select $H_\gamma$ as the representative Balmer line, since it has the largest amplitude of variation among the Balmer lines. Then we average over the Mg$_1$, Mg$_2$, and Mg$_b$ as the composed line index of Mg. We also average over all 9 iron lines as the composed line index of Fe. Finally, we select G band (CH) and TiO2 to represent for the molecular bands. In total, we give 5 (composed) line indices for all selected stars. 

Although CaII K line is frequently used in classifications and parameterizations, we decide not to use it because it does not provide extra information about spectral types and is located at around the blue end, in which the wavelength calibration and the efficiency of the instrument are not as good as other lines, making the line index of CaII K be not very stable.

\section{Line indices-based classification}\label{sect:line}
\begin{figure}
 \begin{minipage}{15cm}
 \centering
 \includegraphics[scale=0.55]{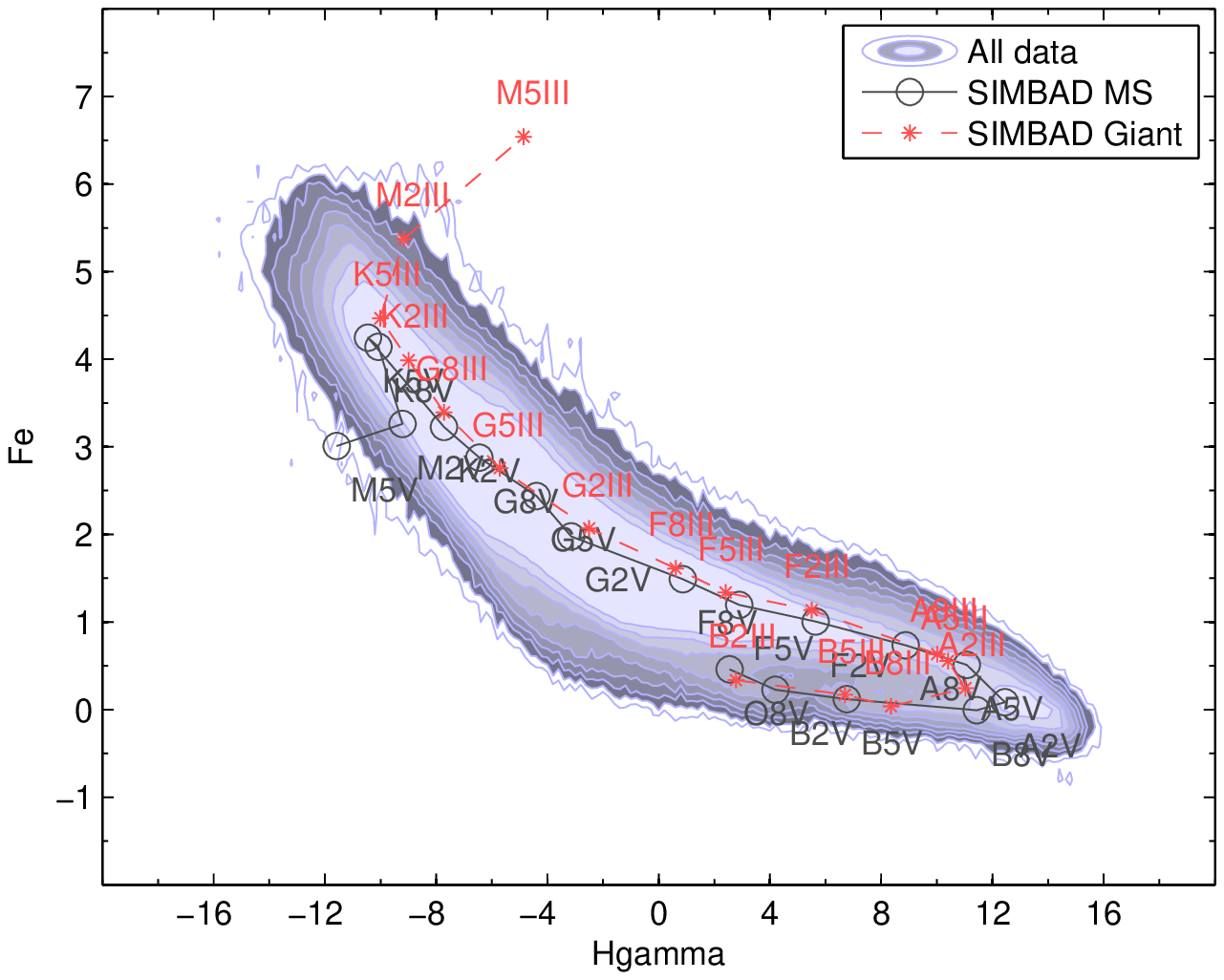}
 \includegraphics[scale=0.55]{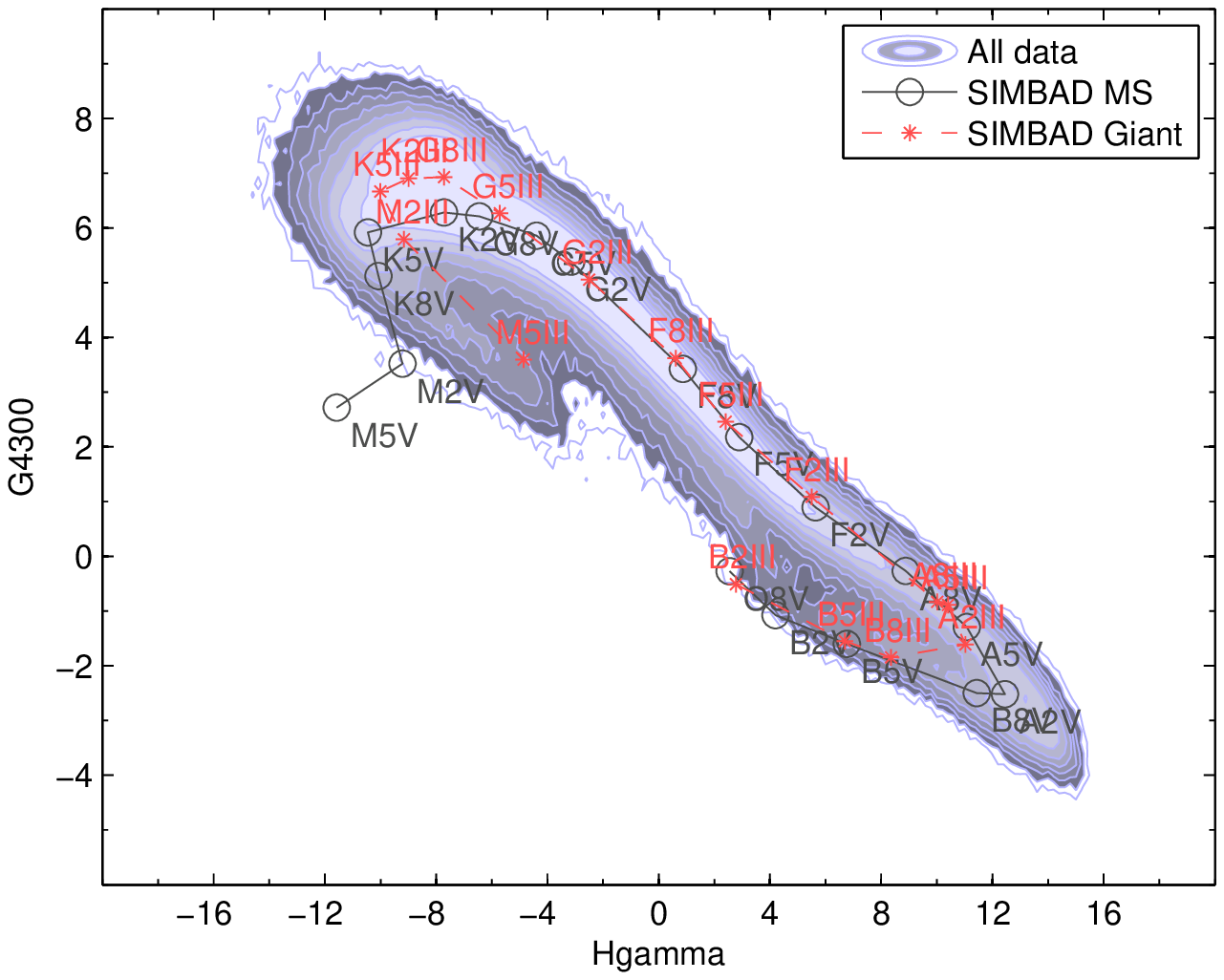}
 \includegraphics[scale=0.55]{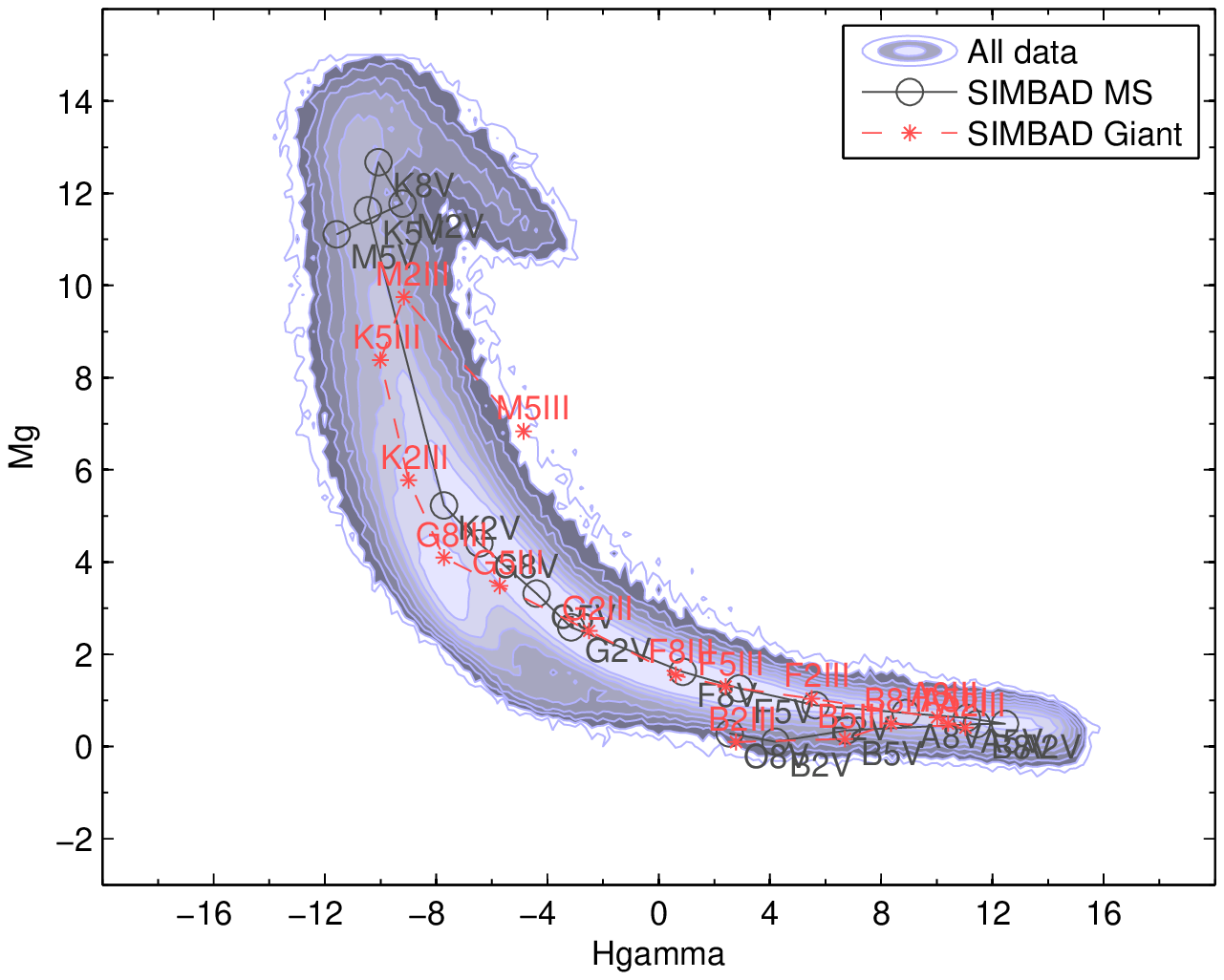}
\includegraphics[scale=0.55]{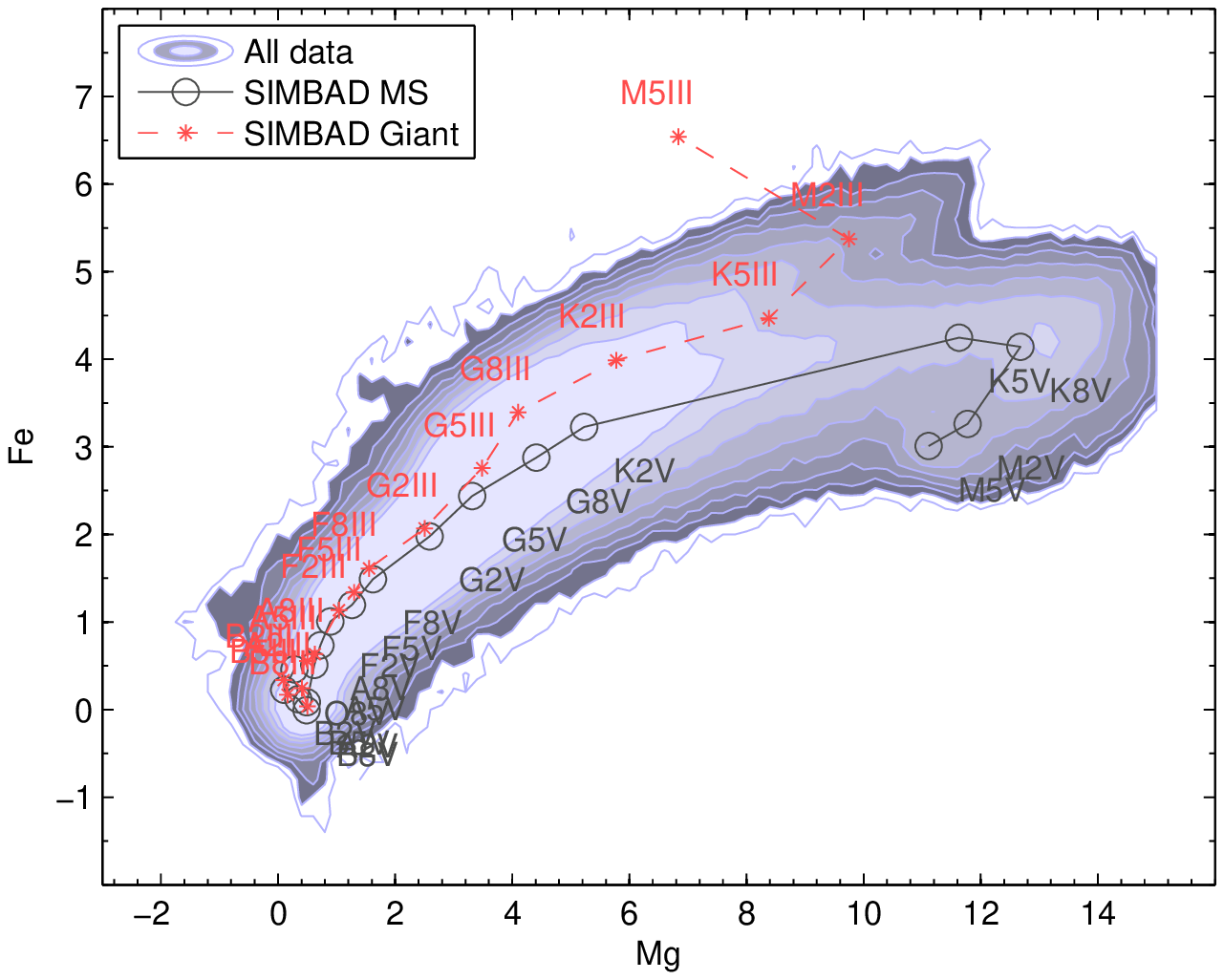}
 \includegraphics[scale=0.55]{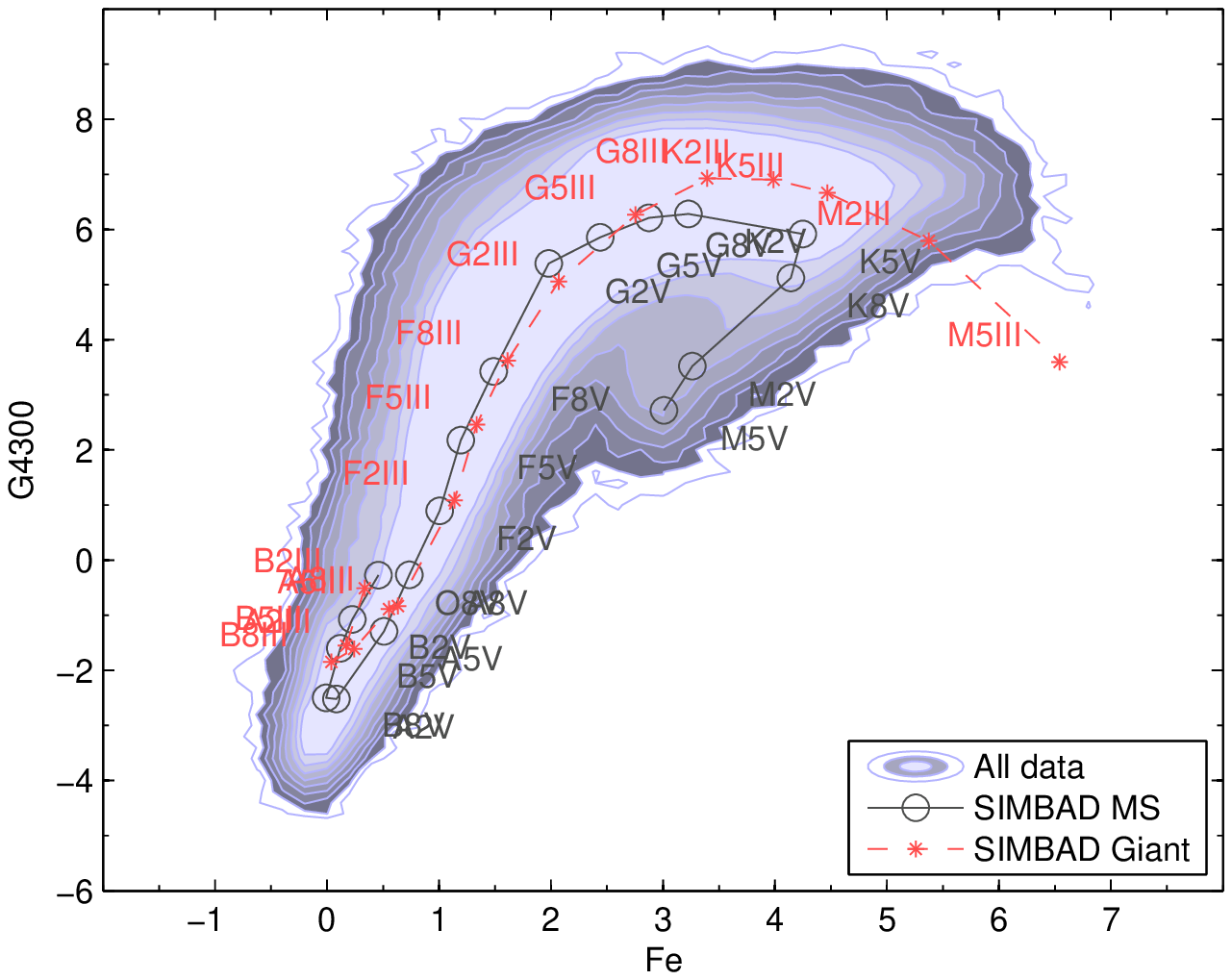}
\includegraphics[scale=0.55]{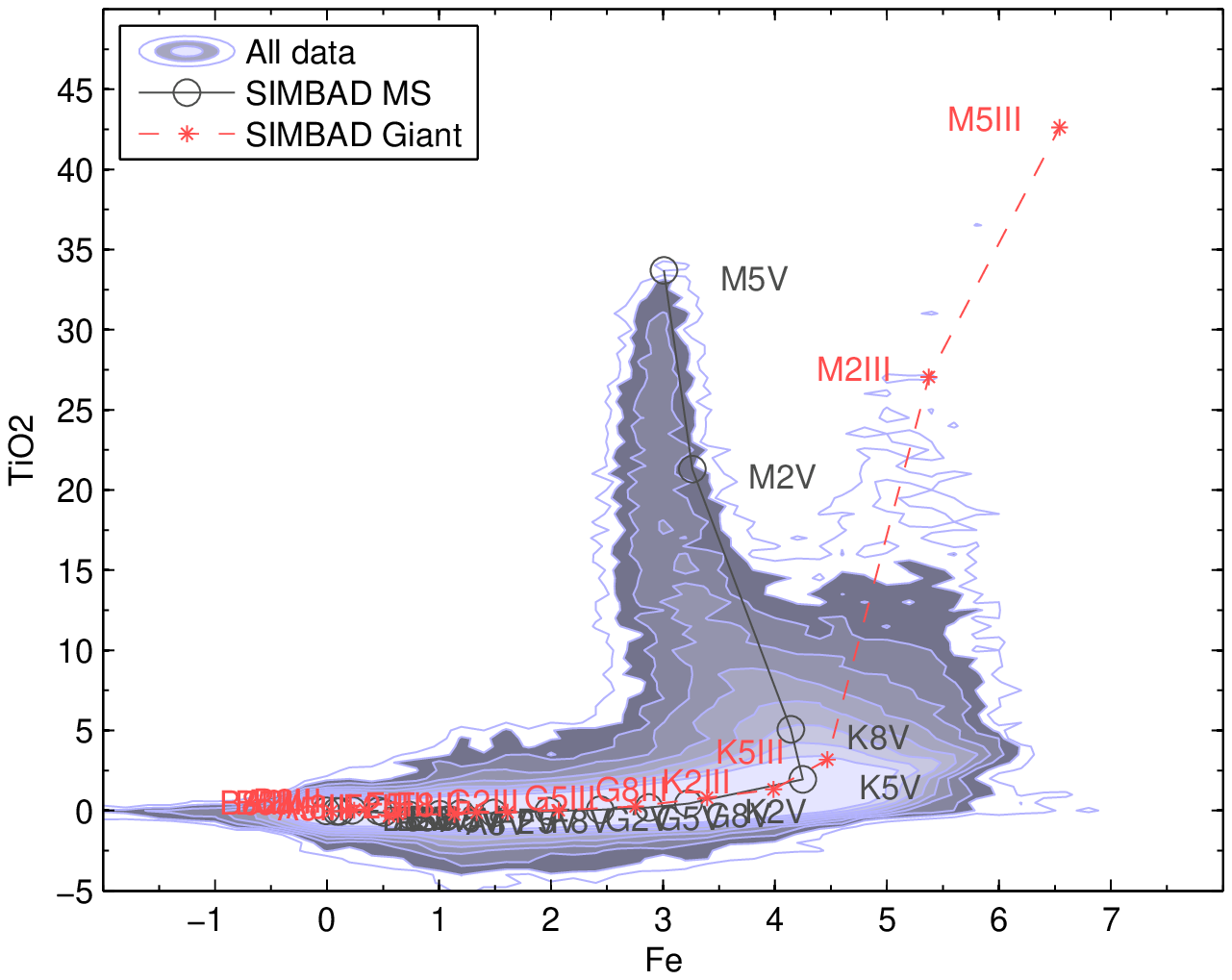}
 \end{minipage}
 \caption{The contours show the distribution of the LAMOST stars in the space of line indices. The top-left, top-right, middle-left, middle-right, bottom-left, and bottom-right panels are for the H$_\gamma$ vs. Fe, H$_\gamma$ vs. G band, H$_\gamma$ vs. Mg, Mg vs. Fe, Fe vs. G band, and Fe vs. TiO2 planes. The dark lines with circles indicate the stellar loci of the main-sequence stars with the MK marks from the SIMBAD database, while the RED dashed lines with asterisks indicate the stellar loci of the giant stars (type IV/III) with the MK marks from the SIMBAD database.}\label{fig:stellarlocusMSGNT}
 \end{figure}

\begin{figure}
 \begin{minipage}{15cm}
 \centering
 \includegraphics[scale=0.55]{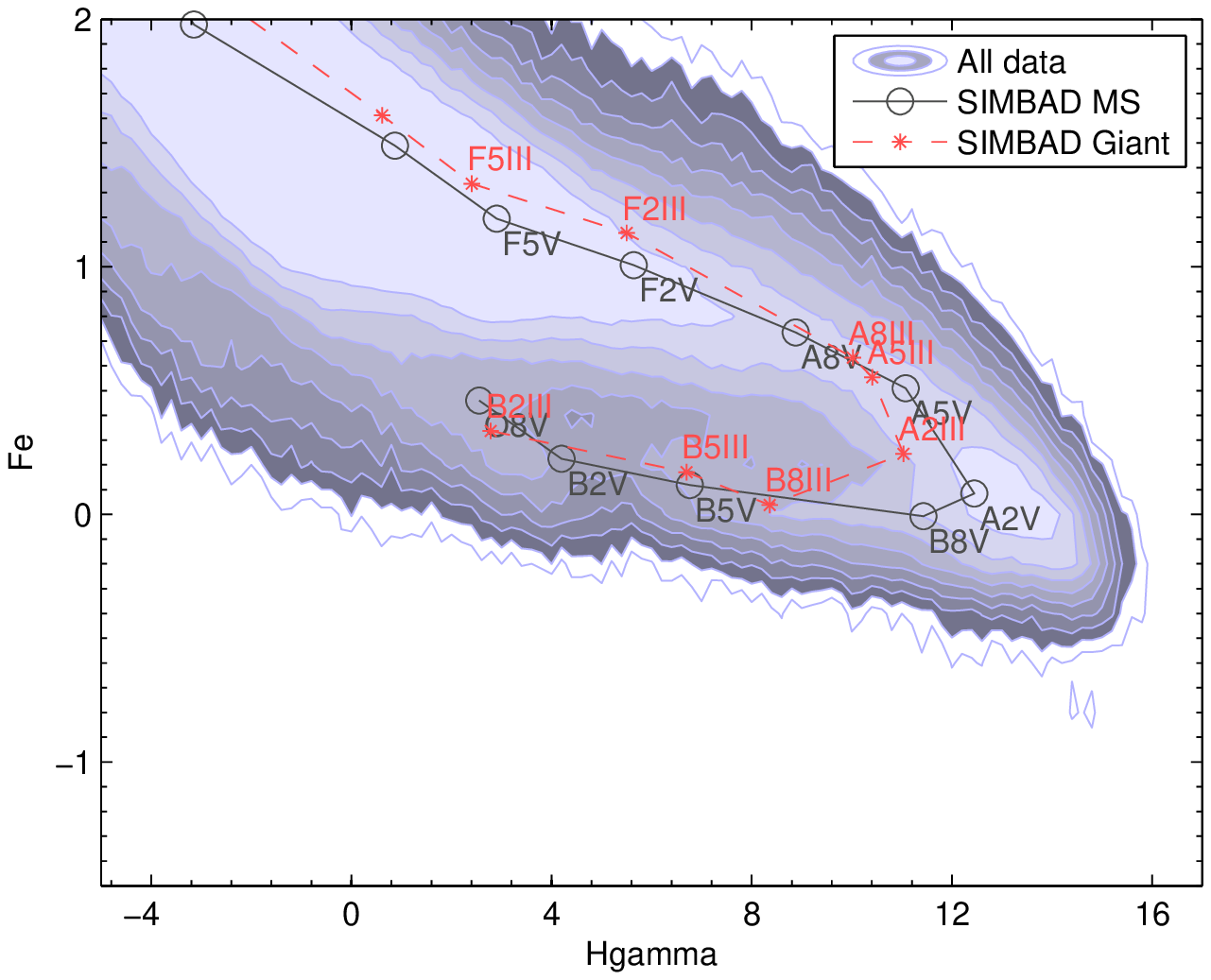}
 \includegraphics[scale=0.55]{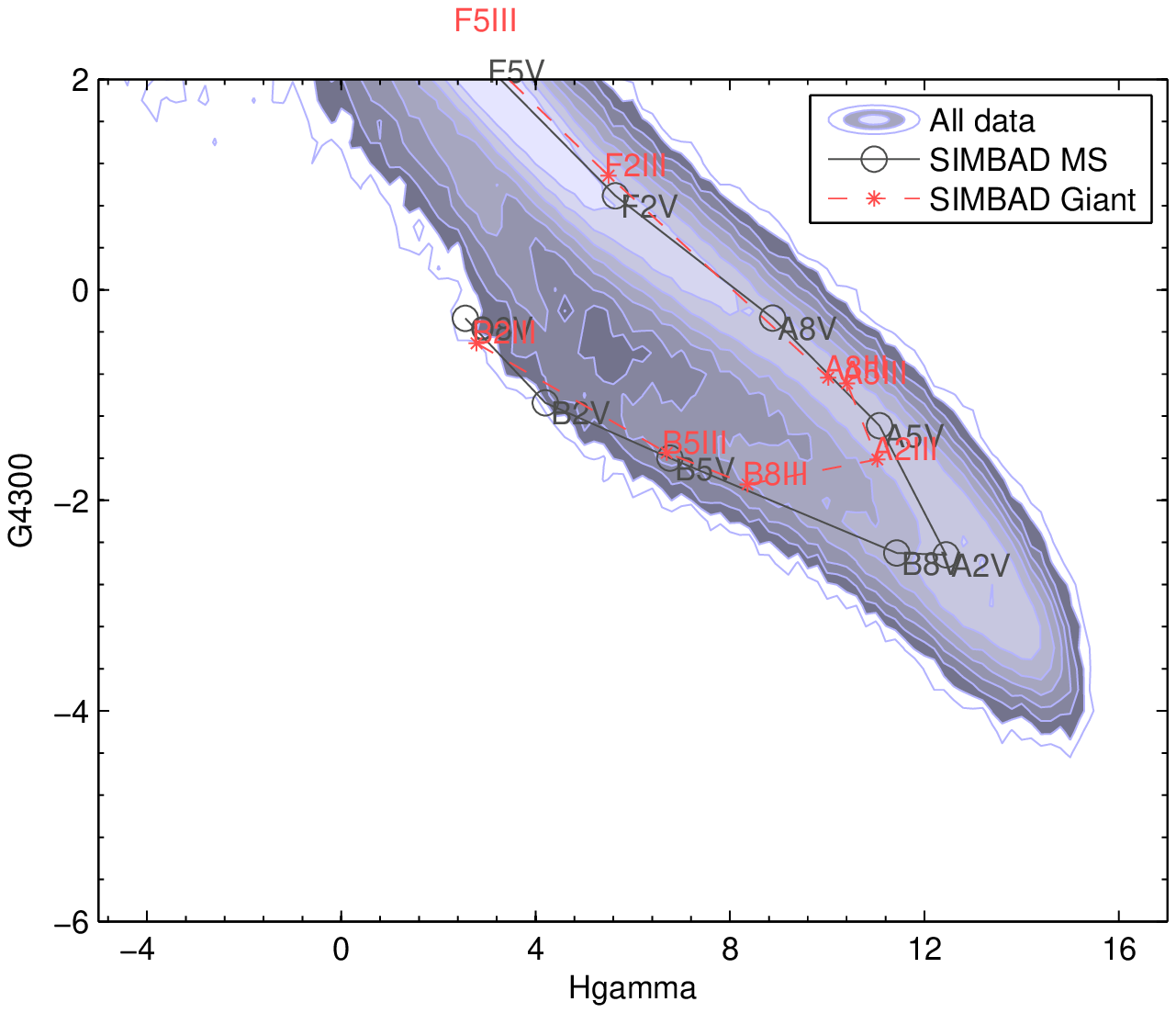}
 \includegraphics[scale=0.55]{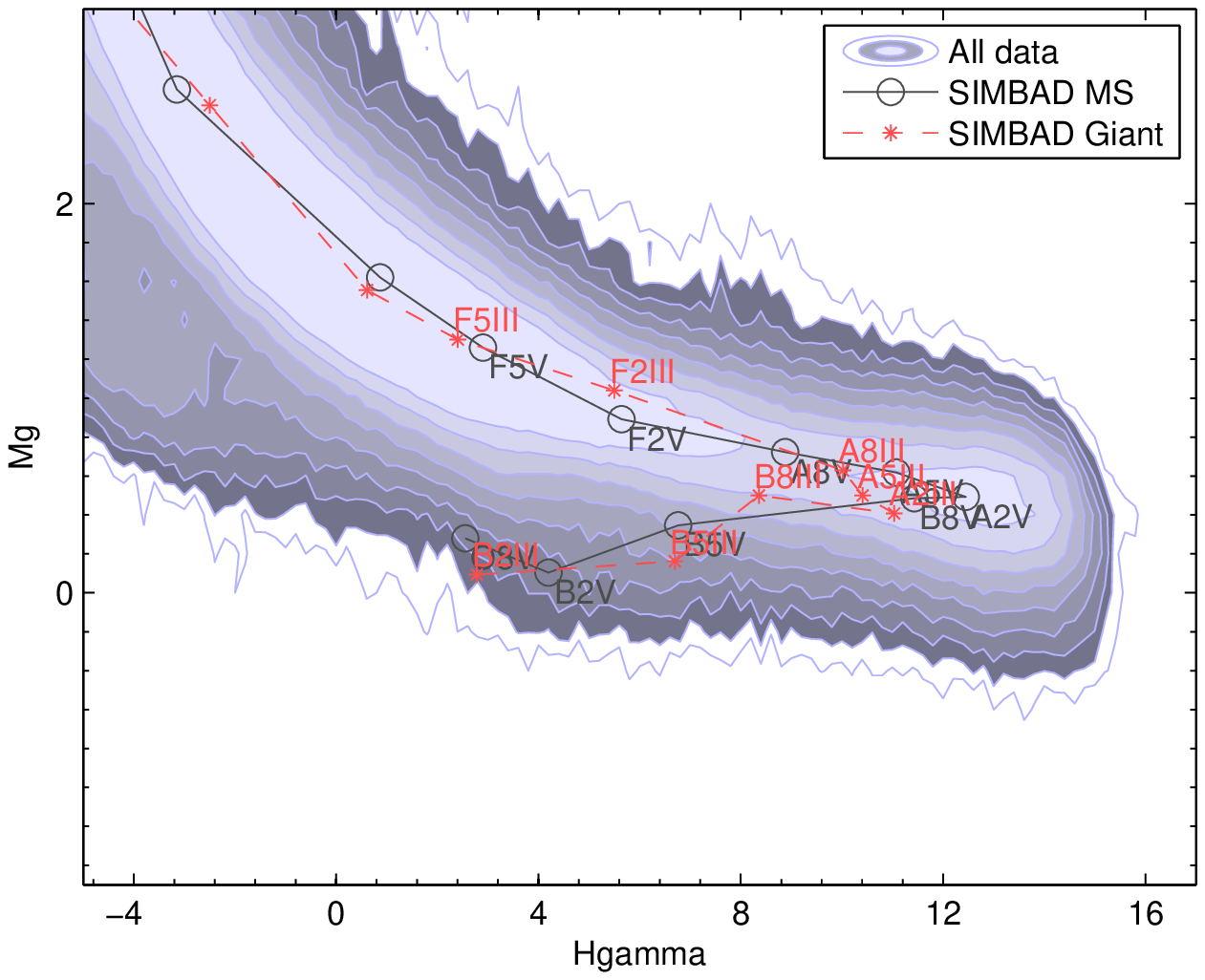}
\includegraphics[scale=0.55]{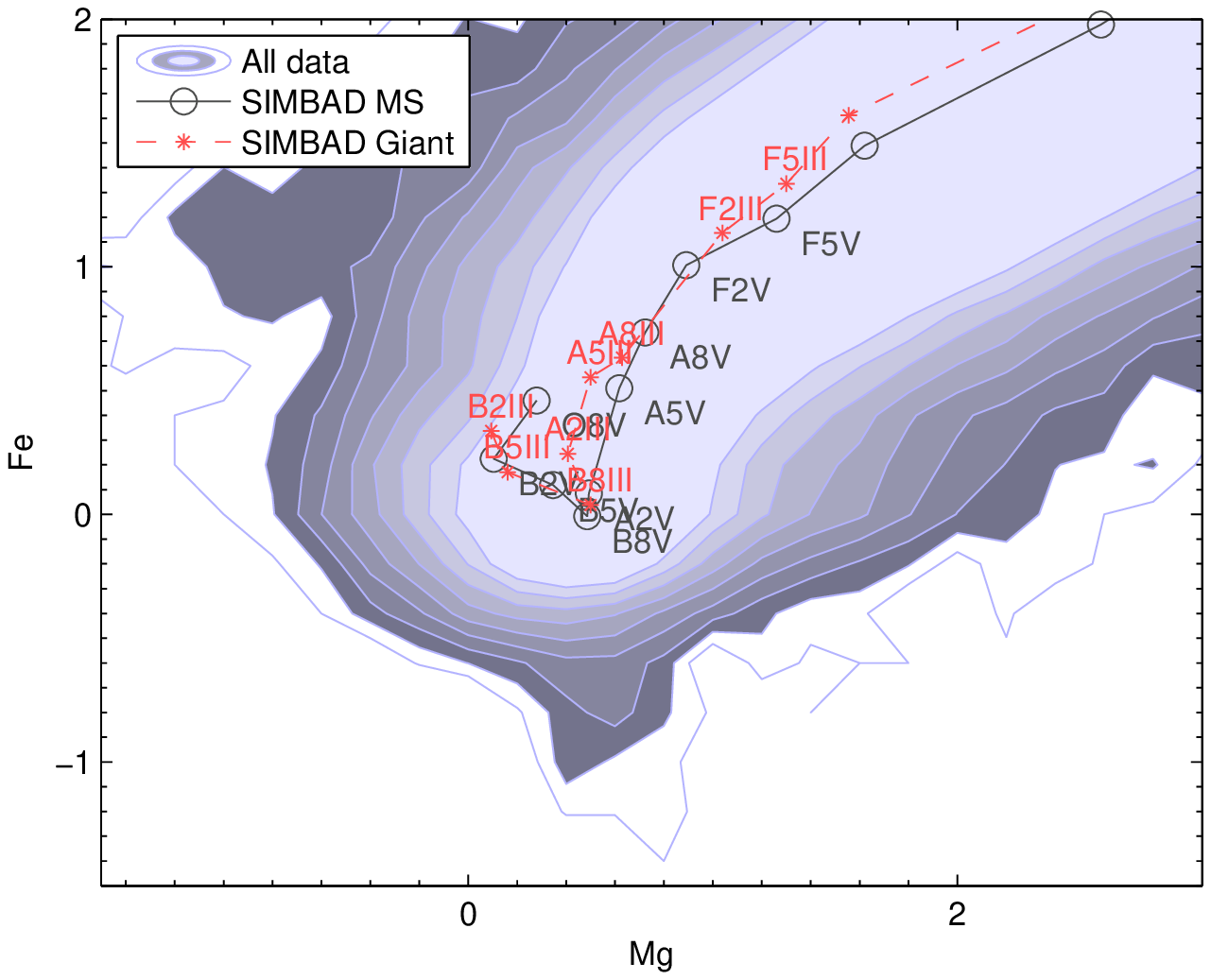}
 \includegraphics[scale=0.55]{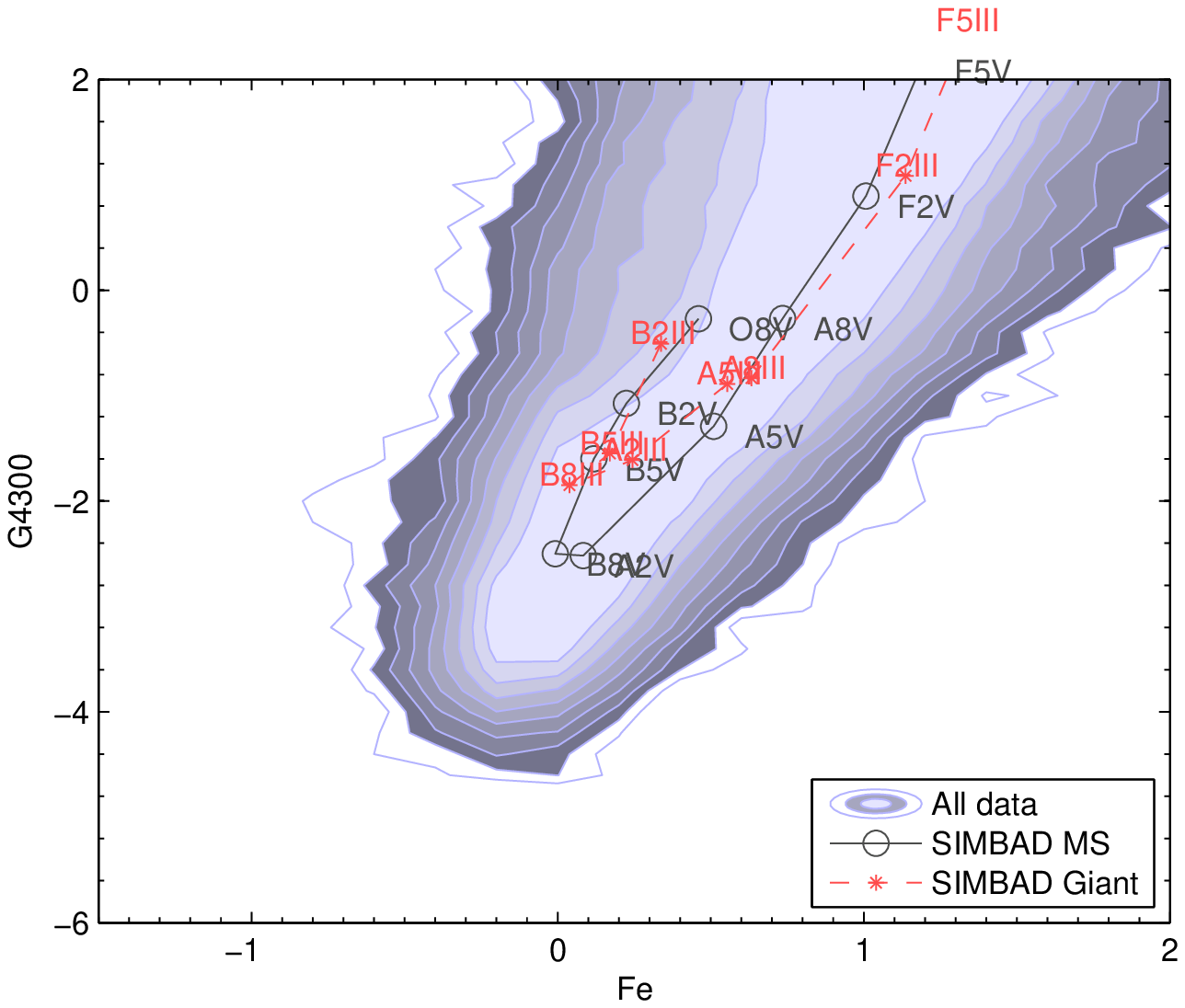}
\includegraphics[scale=0.55]{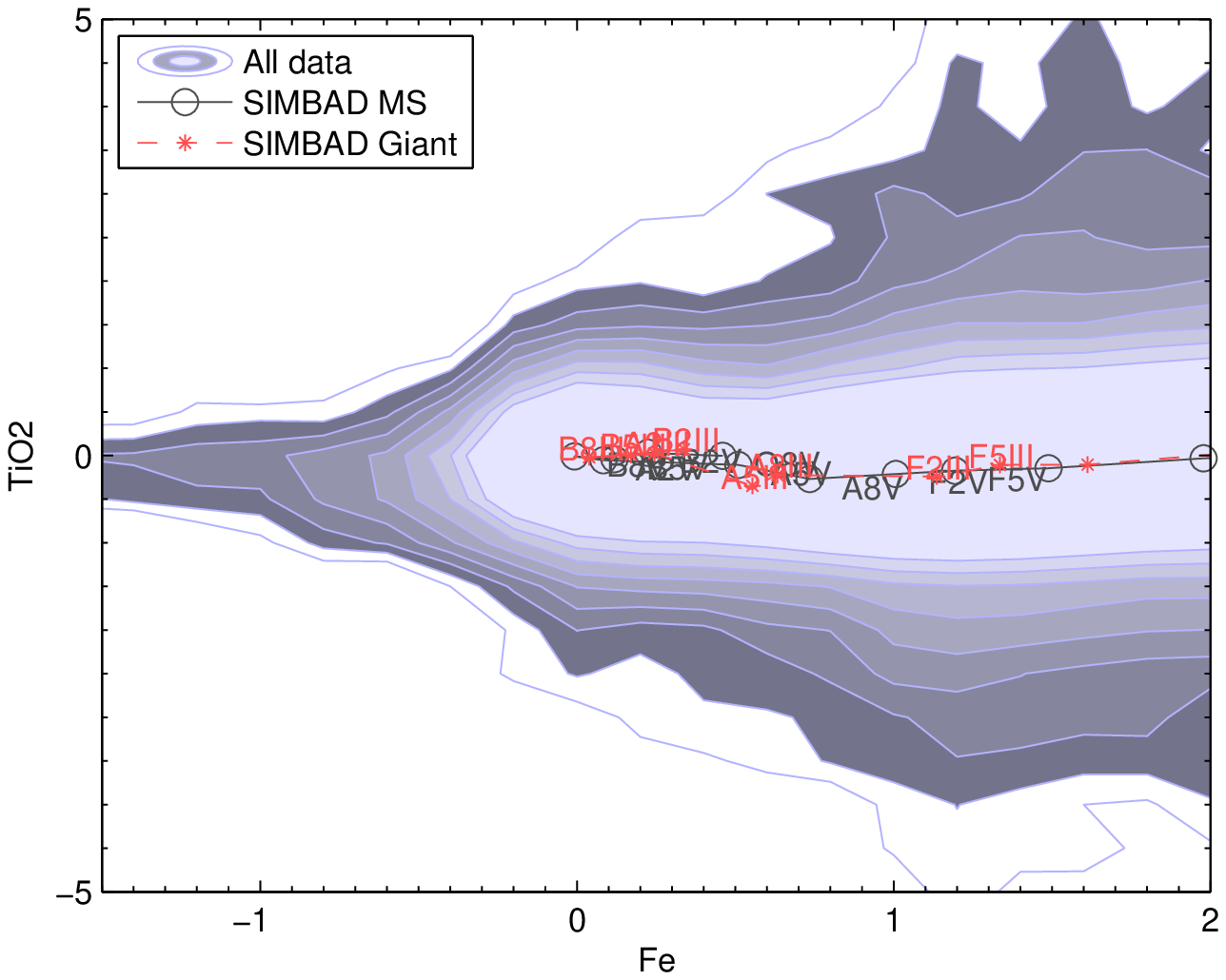}
 \end{minipage}
 \caption{The panels are same as their counterparts in Fig~\ref{fig:stellarlocusMSGNT}, but are zoomed in to the details around the locus of the early type stars}\label{fig:stellarlocusMSGNTzoom}
 \end{figure}

Fig~\ref{fig:stellarlocusMSGNT} shows the stellar loci in the space of the 5 line indices, H$_\gamma$, Mg, Fe, G band, and TiO2 for all 1.5 million selected stars (only show their distributions in blue contours). The unit of the x- and y-axes is $\rm\AA$. The hollow circles with the neighboring dark gray labels mark the median positions of the main-sequence MK classes from the SIMBAD catalog. For instance, a hollow circle with a label ``G2V" is the median value of the stars with the type G0V, G1V, G2V, and G3V. And a symbol with ``G5V" is the median of all G3V, G4V, G5V, and G6V stars. The neighboring circles overlap with each other by one decimal subtype in order to make the stellar locus smoother. Similarly, the red asterisks with the neighboring labels indicate the locus of the giant star MK classes. The detailed positions of these circles (asterisks) for main-sequence (giant) stars are listed in Table~\ref{tab:MSlocus} (Table~\ref{tab:giantlocus}). Fig~\ref{fig:stellarlocusMSGNTzoom} zooms in to the smaller regions for better illustration of the early type stars.

First, the stars from O to M type can be well separated and ordered in H$_\gamma$ vs. G4300 plane, shown in the top-right panel of Fig~\ref{fig:stellarlocusMSGNT}. In H$_\gamma$ vs. Fe plane (top-left panel), the stars from O to G type are well separated, while the M and K type stars overlap at the top of the stellar loci and hard to be disentangled. Similar trend is shown in H$_\gamma$ vs. Mg plane in the middle-left panel of Fig~\ref{fig:stellarlocusMSGNT}. However, the late type main-sequence stars are well disentangled in Mg vs. Fe, Fe vs. G4300, and Fe vs. TiO2 planes shown in the middle-right, bottom-left, and bottom-right panels, respectively, while the early type stars in these planes are clumpy and hard to be separated from each other. Combined with the 5 line indices, we are able to separate all types of main-sequence stars from O to M type. 

Second, the separation of the luminosity type works well for K and M giant stars. Especially in Mg vs. Fe (middle-right panel of Fig~\ref{fig:stellarlocusMSGNT}) and Fe vs. TiO2 (bottom-right panel) planes, the cool star ends of the stellar loci of main-sequence and giant stars go to different directions. In Mg vs. Fe plane, the locus of the main-sequence stars goes down toward smaller Fe and larger Mg indices in the coolest end, while the locus of the giant stars goes up toward larger Fe but smaller Mg indices. Similar trend can also be seen in Fe vs. TiO2 plane. However, It is very hard to disentangle the early type giant stars, e.g. B, A, and F type giant stars. These types of giant stars are located at almost exactly the same position as the same type main-sequence stars. According to Gray \& Corbally~(\cite{gray09}), some weaker lines, such as OII (at 4070, 4076, 4348, and 4416$\rm\AA$), SIIV at 4116$\rm\AA$ etc., may be helpful to discriminate the luminosity types for B type stars. However, they are very weak in the low-resolution LAMOST spectra and may be significantly affected by the noise. 

It is worthy to point out that the variation of Fe index for the late type stars are mostly not exactly related with the Fe lines, but significantly affected by the prominent molecular bands, e.g. TiO, happened to overlap at the same wavelength. And the response of the Mg index to the cool stars are actually dominated by the MgH band.  

{It is also noted that the dispersions shown in Fig~\ref{fig:stellarlocusMSGNT} are not only contributed by the uncertainties of the line indices, which is only about 0.1$\rm\AA$. The dispersions may be intrinsic and related with the broad diversity of the metallicity. In this paper, we mainly focus on the effective temperature, which corresponds to the spectral types, and the surface gravity, which is related with the luminosity. The effect of the metallicity in spectral classification may be more complicated, because it also reflects the evolution of the different stellar populations. We would like to leave this topic in future works. }

The classification of the stars based on the stellar loci can be done by looking up the lines indices in Tables~\ref{tab:MSlocus} and~\ref{tab:giantlocus}. For any statistical study of the Milky Way, one can conveniently select stars located in a segment of the stellar loci in Figs~\ref{fig:stellarlocusMSGNT} and~\ref{fig:stellarlocusMSGNTzoom} according to the marked MK classes. Compared with the classical MK classes, the stellar loci in line indices, acting just like the color indices in a multi-band photometric system, provide natural and continuous sequences of the stars, which are easier in quantitative statistics. More discussions can be found in section~\ref{sect:disc}. 
 
\section{SVM-based classification}\label{sect:svm}
Alternatively, we can also translate the line indices-based stellar loci into MK class system for individual stars. To do this, we employ a SVM algorithm to automatically assign the proper MK class to a stellar spectrum.

SVM is a supervised machine learning algorithm for classification and regression (Cortes et al.~\cite{cortes95}). In general, a supervised algorithm uses a small sample with the multi-dimensional input variables and known labels of classes as the training dataset. The SVM classification is built in two steps. First, with the training dataset, the optimized non-linear boundaries among different classes in the input space is determined and defined by a subset of the training dataset, which is called the support vectors, located around the boundaries. Second, for a given input data, the trained SVM model gives a prediction of the class depending on where the input data is located with respect to the support vectors. A typical sample of SVM classification can be found in Liu et al.~(\cite{liu14a}) and a sample of SVM regression can be found in Liu et al.~(\cite{liu12}) and Liu et al.~(\cite{liu14b}).

Chang \& Lin~(\cite{libsvm}) provides a multi-programming language package, LIBSVM\footnote{http://www.csie.ntu.edu.tw/$\sim$cjlin/libsvm/}, to implement the SVM algorithm. Here, we use LIBSVM to classify the stars into MK types based on the line indices. We arbitrarily separate the 3,134 stars with both high signal-to-noise ratio LAMOST spectra and SIMBAD MK types into two equal-size groups. One group is selected as the training dataset to train the SVM, and the other is used as the test dataset to assess the performance. We use all 27 line indices listed in Table~\ref{tab:lineindices} as the input vector. We only adopt 6 classes, which are OB, A, F, G, K, and M, and ignore the decimal subtypes and the luminosity types in the SVM classification. O and B types are merged as one class since there are only very few O type stars in the sample. 

\begin{figure}
 \begin{minipage}{15cm}
 \centering
 \includegraphics[scale=0.5]{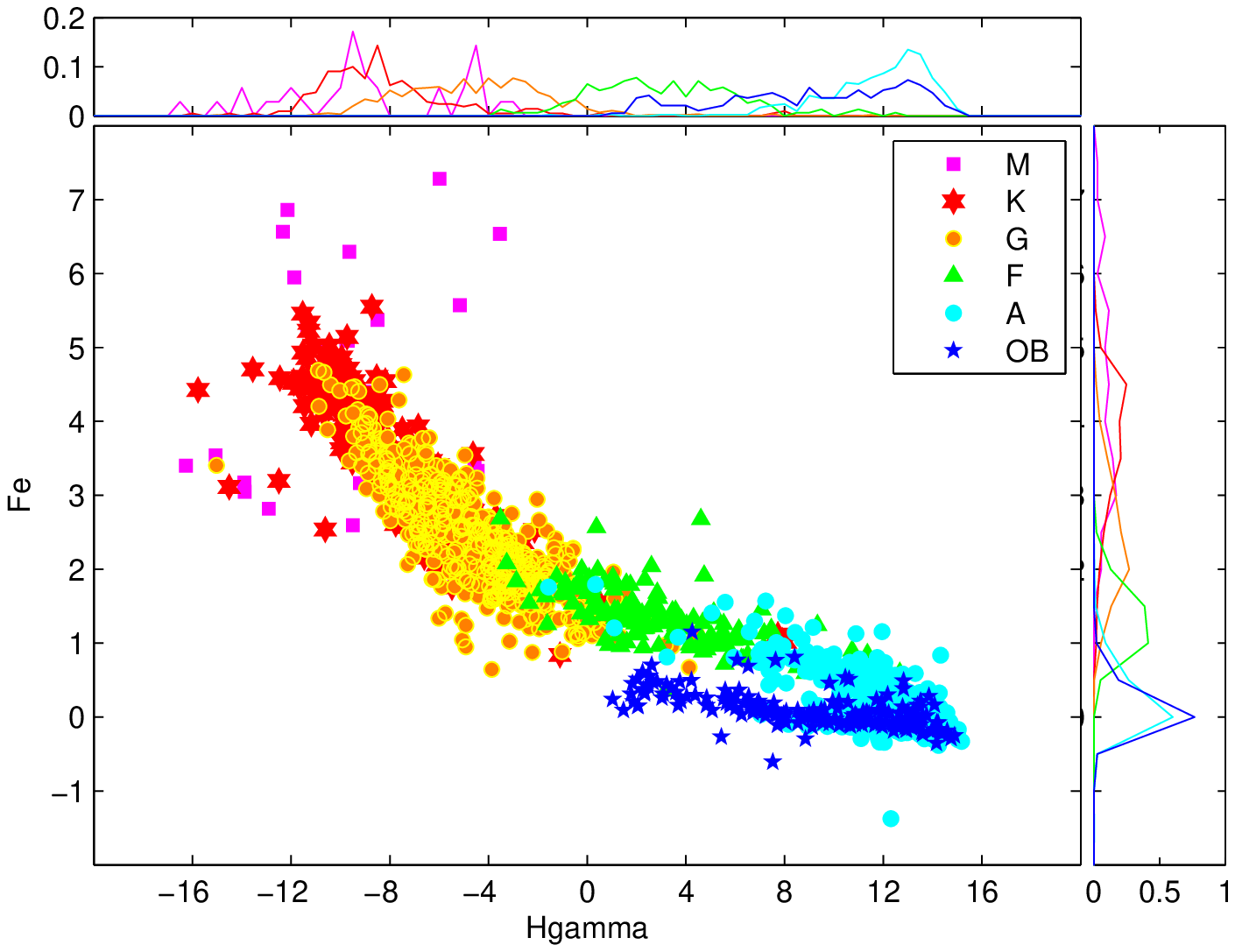}
 \includegraphics[scale=0.5]{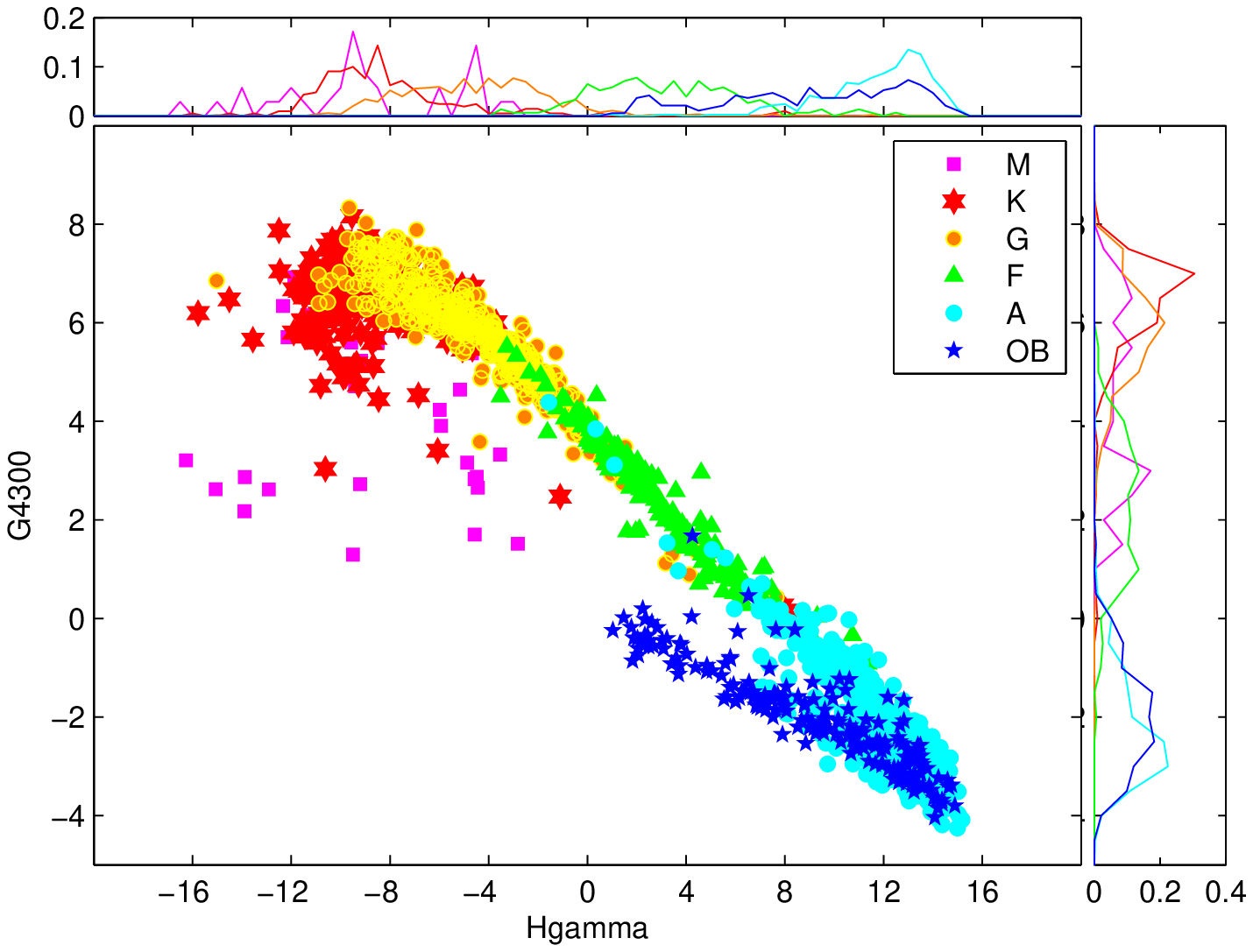}
 \includegraphics[scale=0.5]{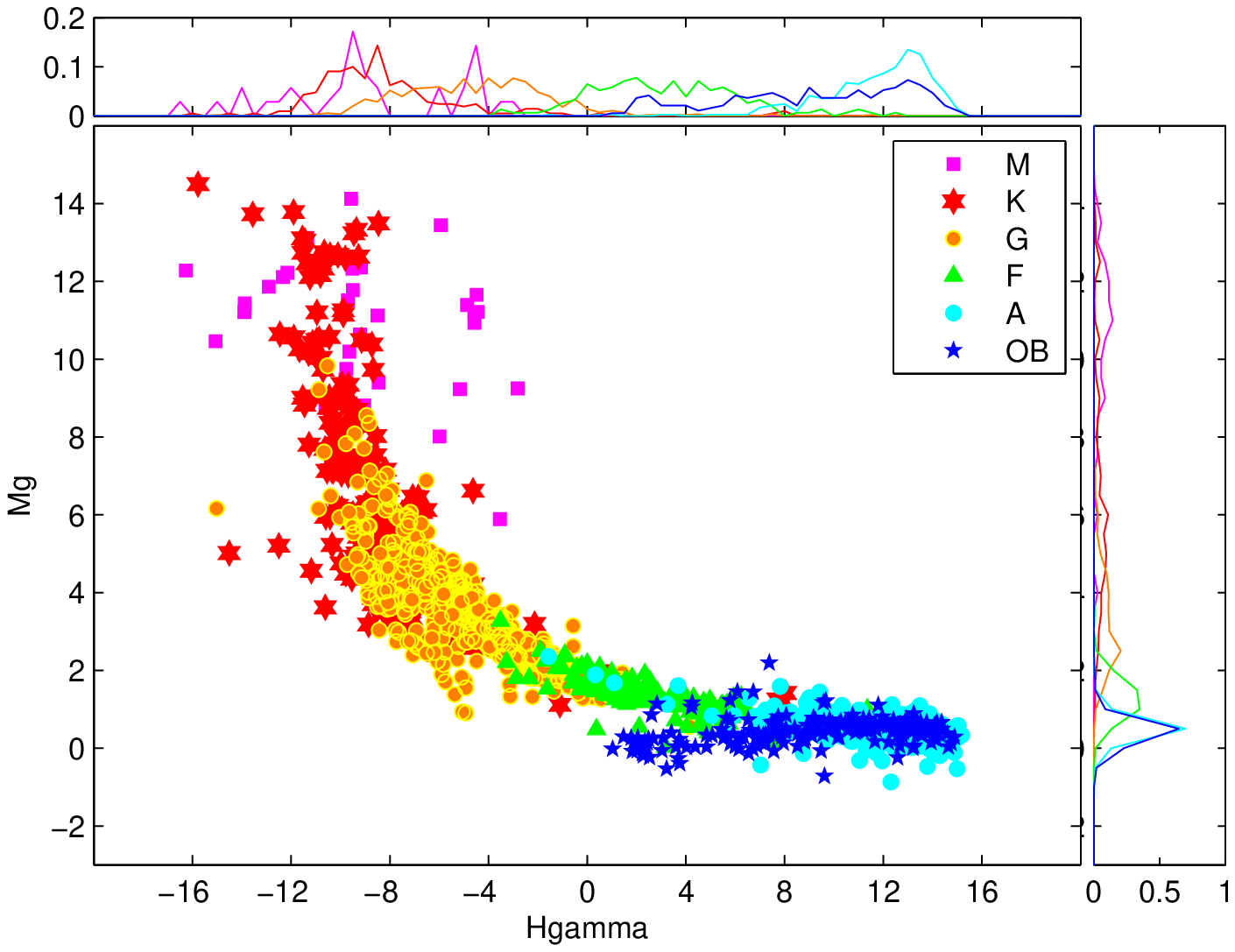}
\includegraphics[scale=0.5]{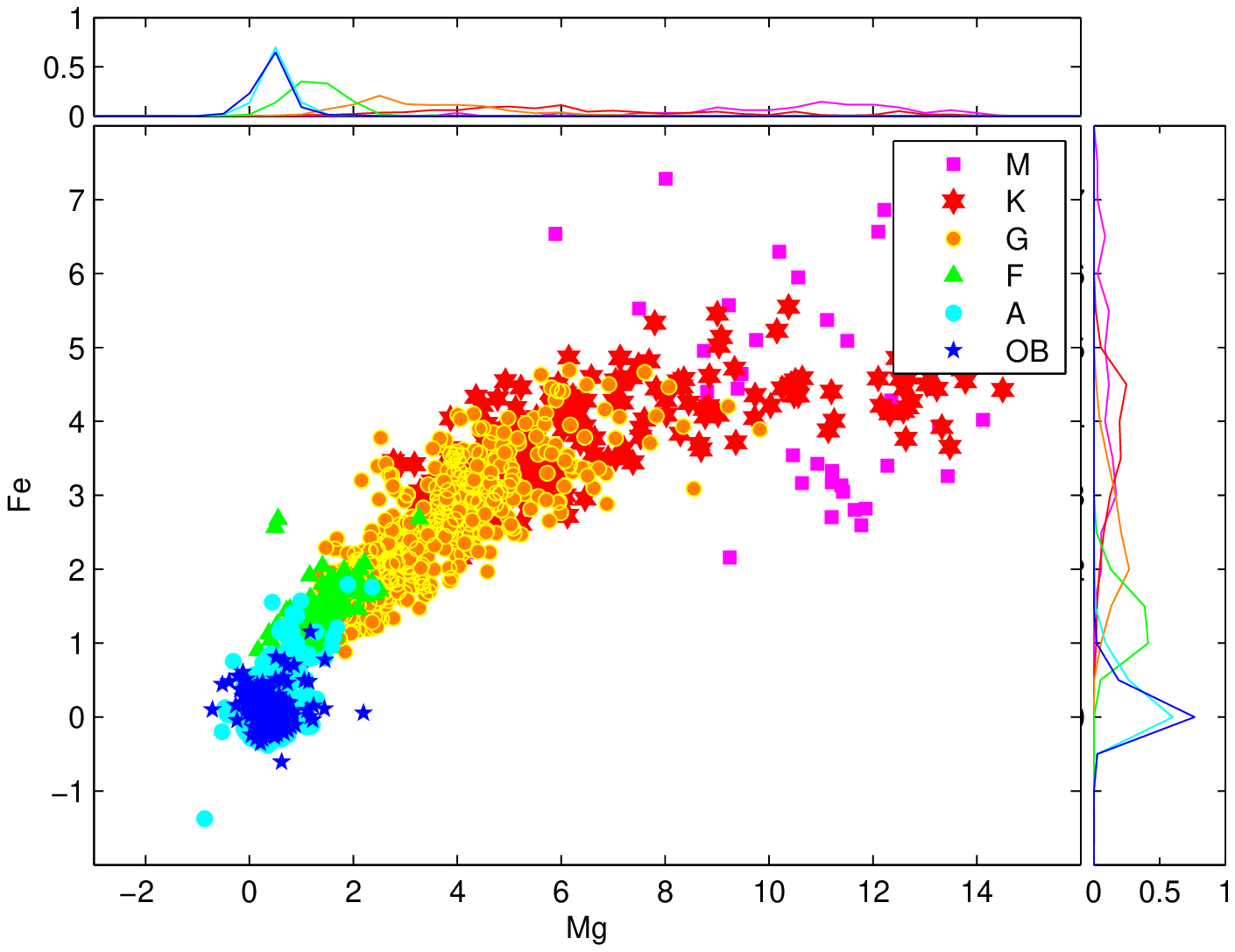}
 \includegraphics[scale=0.5]{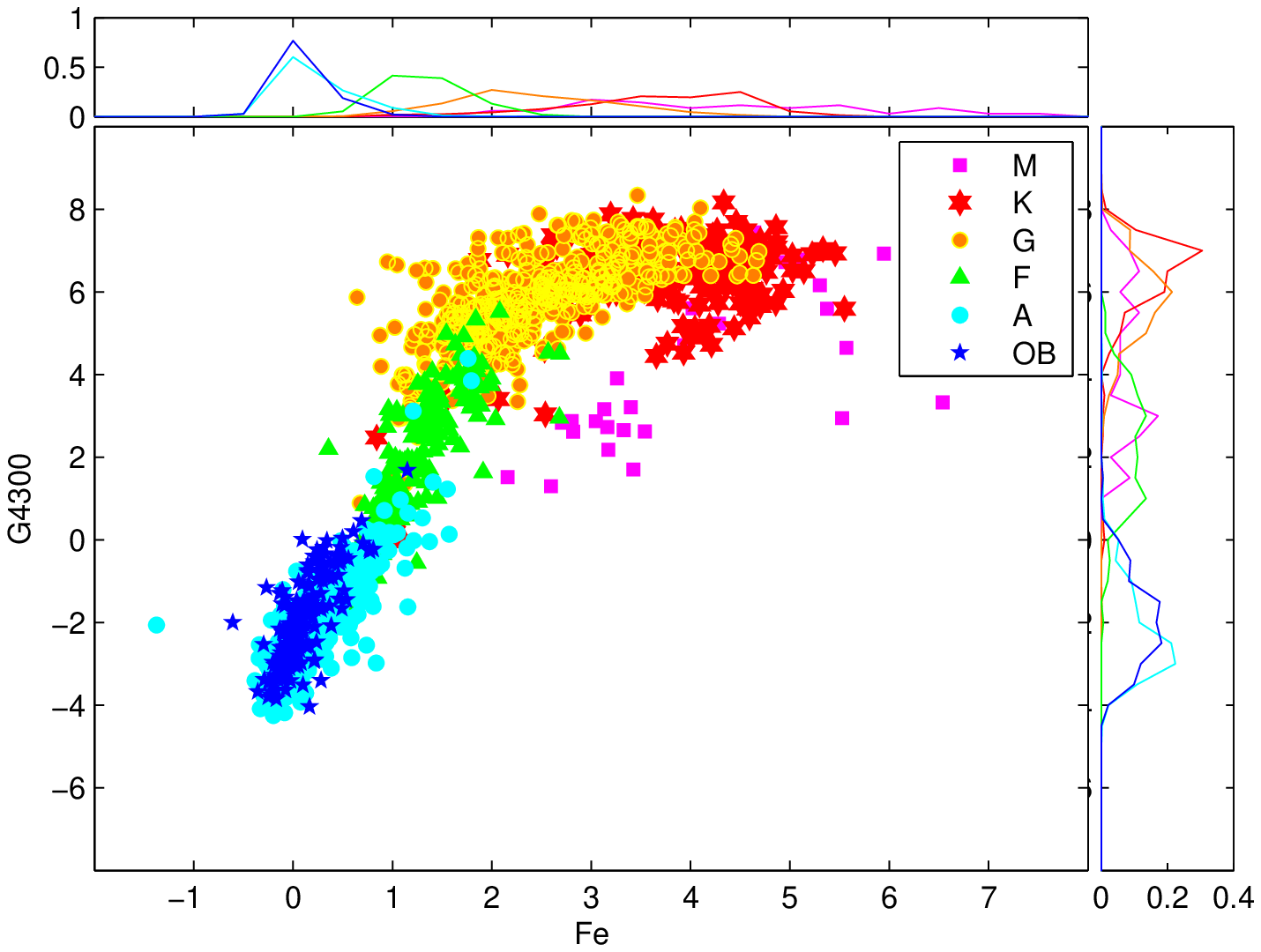}
\includegraphics[scale=0.5]{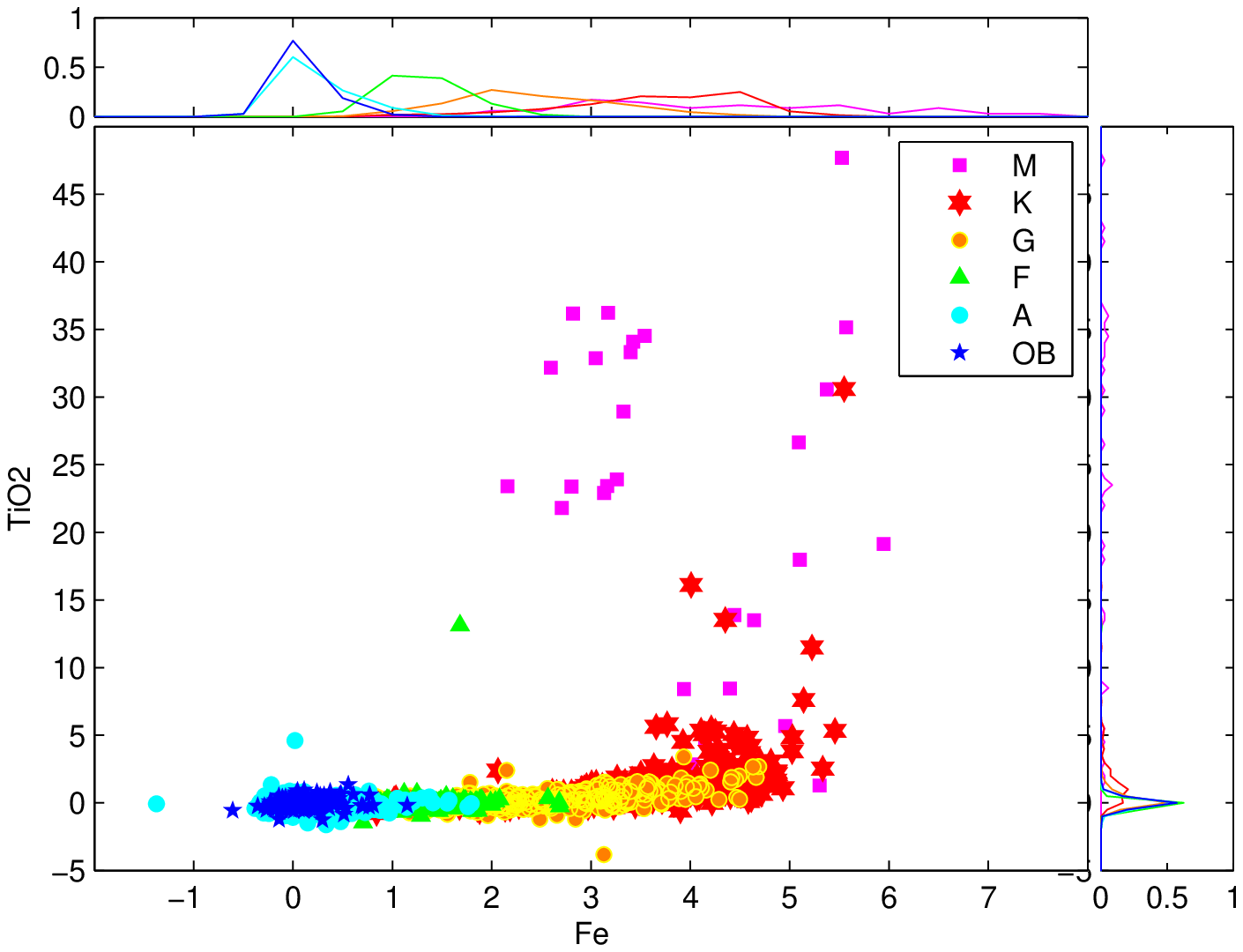}
 \end{minipage}
 \caption{The distribution of the SIMBAD MK classes of the test data for SVM in the space of line indices. The top-left, top-right, middle-left, middle-right, bottom-left, and bottom-right panels show the distributions in H$_\gamma$ vs. Fe, H$_\gamma$ vs. G band, H$_\gamma$ vs. Mg, Mg vs. Fe, Fe vs. G band, and Fe vs. TiO2 planes. The marginalized distributions of one single line index for different spectral types are shown at the right and top edges of each panel. The colors and symbols codes the OB (blue pentagons), A (cyan large circles), F (green triangles), G (orange small circles), K (red hexagons), and M (Magenta rectangles) types.}\label{fig:MKtrue}
 \end{figure}

\begin{figure}
 \begin{minipage}{15cm}
 \centering
 \includegraphics[scale=0.5]{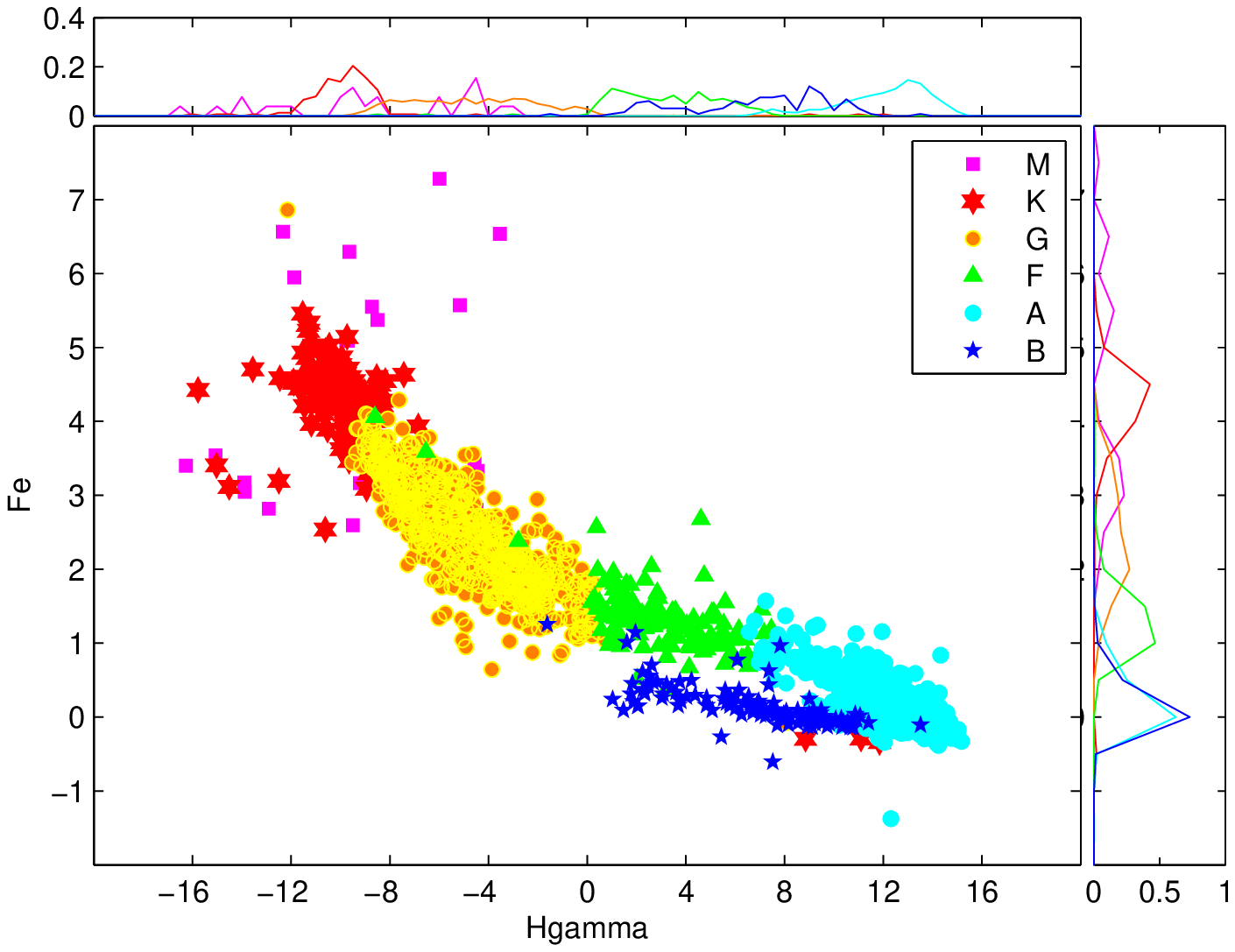}
 \includegraphics[scale=0.5]{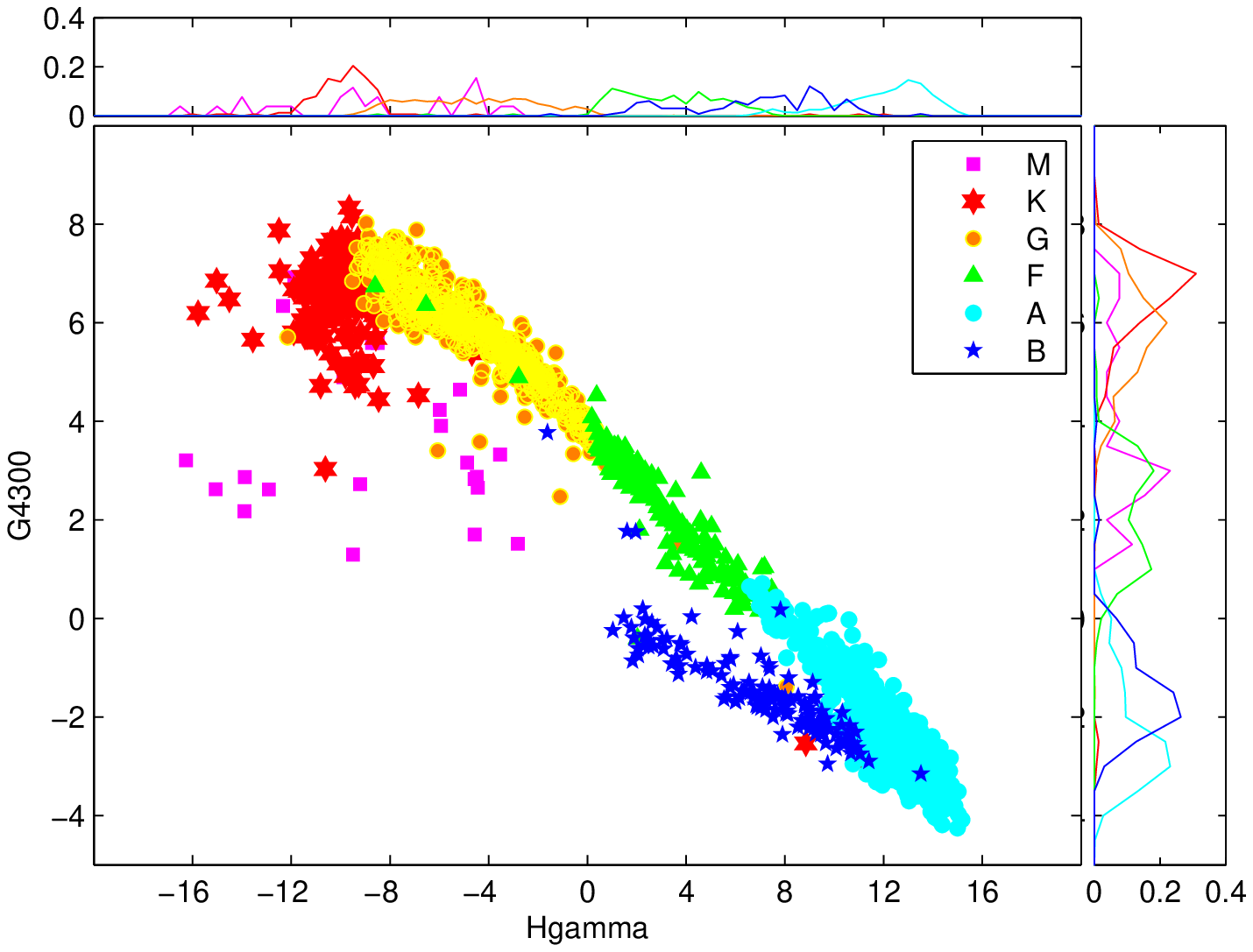}
 \includegraphics[scale=0.5]{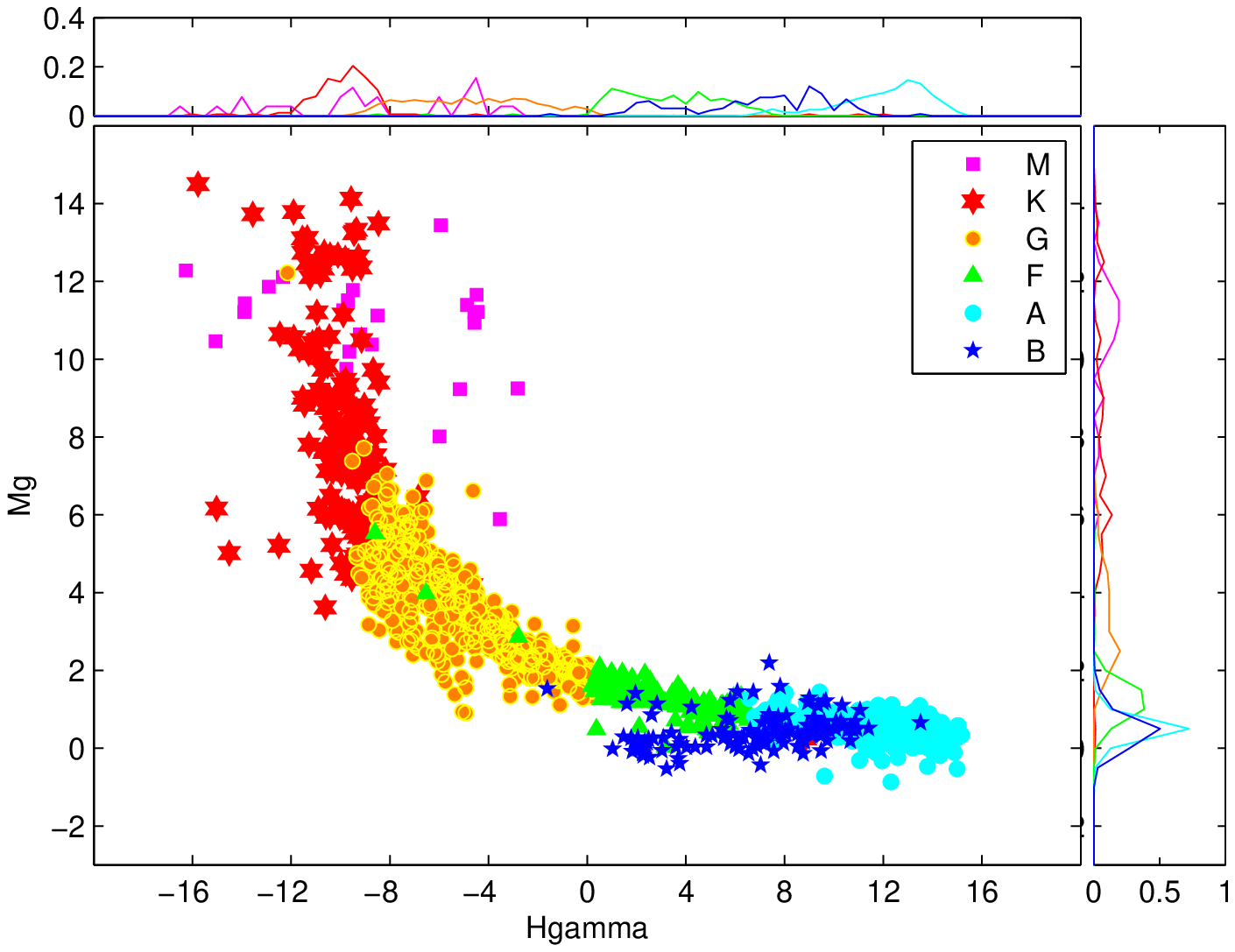}
 \includegraphics[scale=0.5]{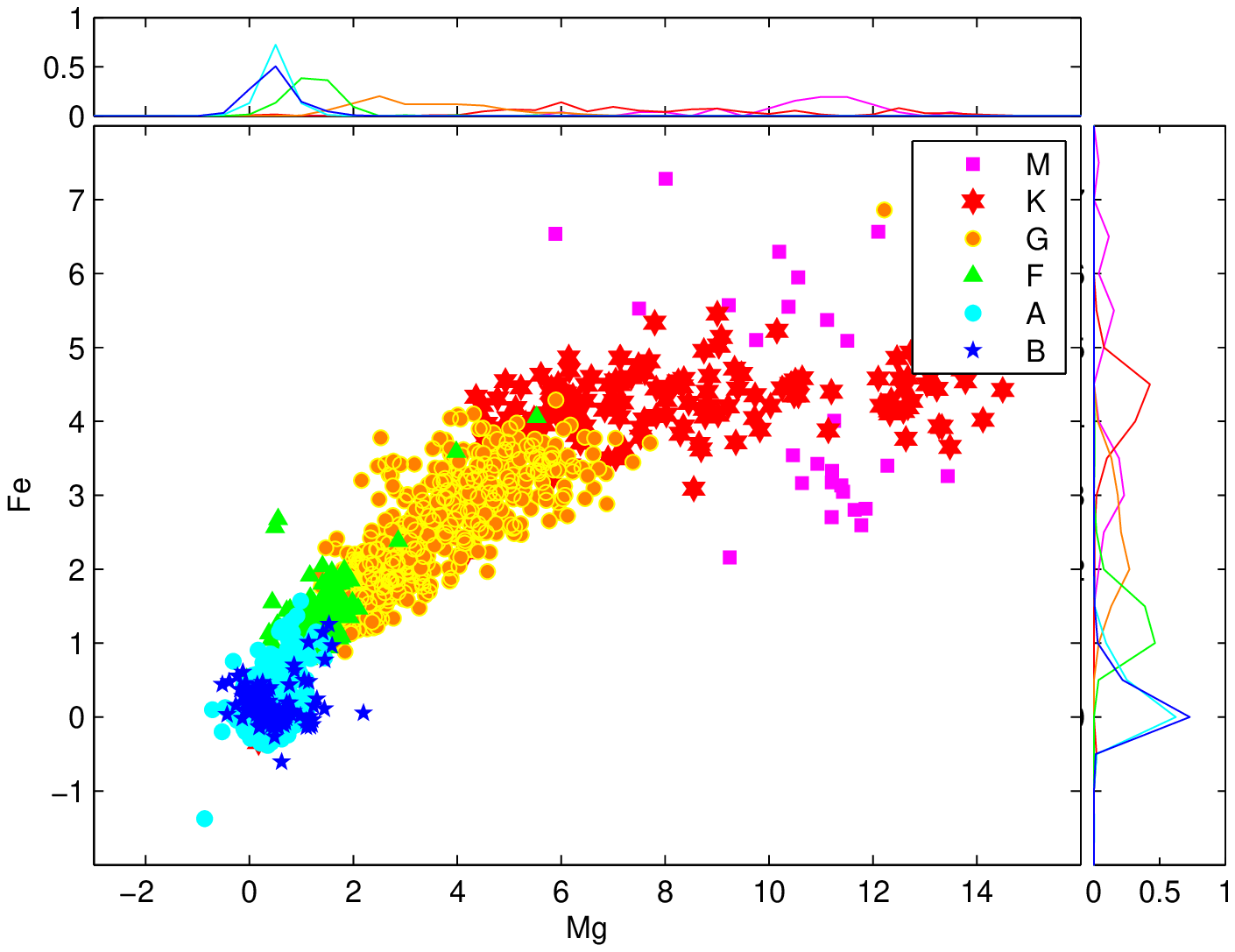}
 \includegraphics[scale=0.5]{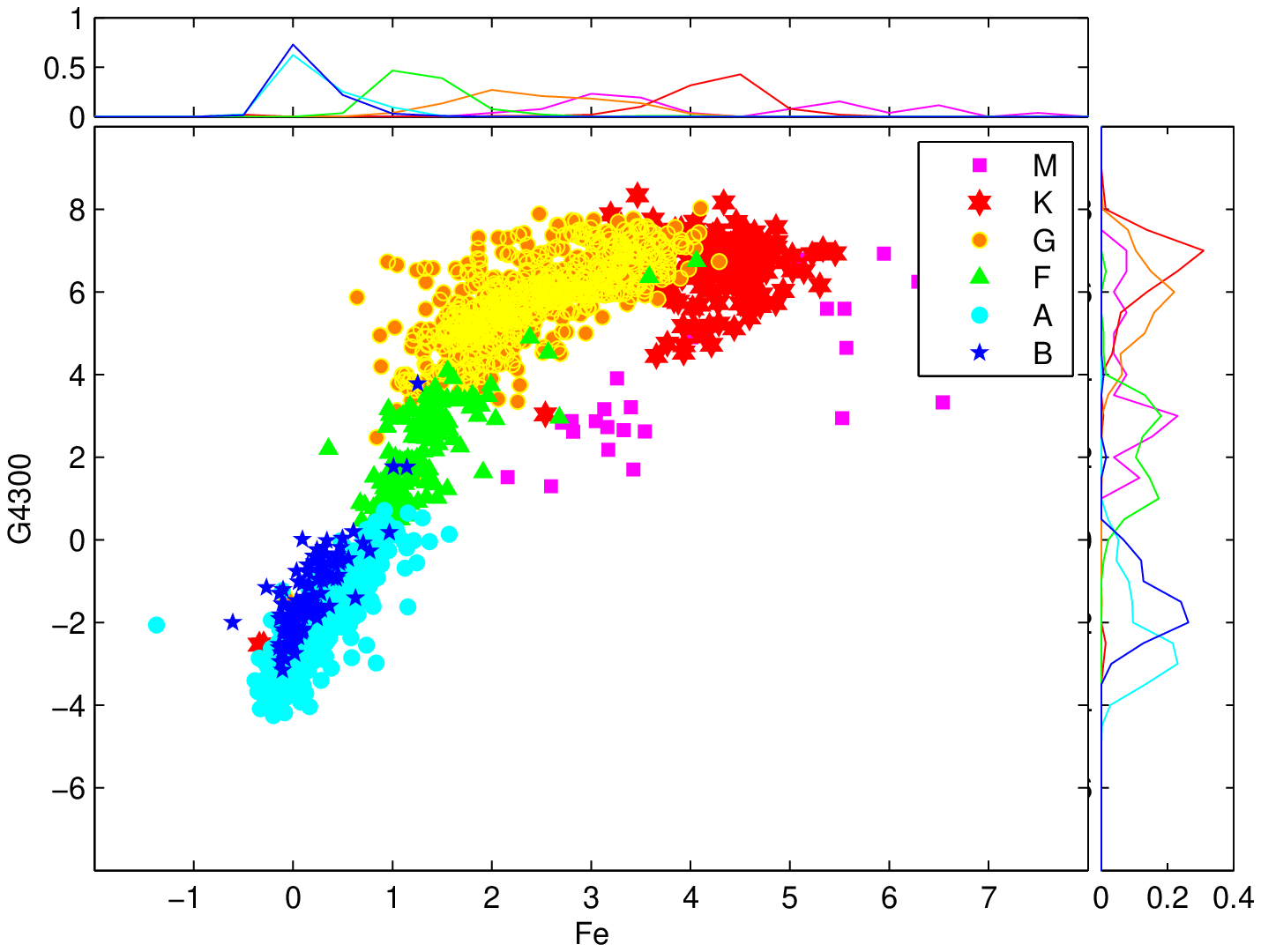}
\includegraphics[scale=0.5]{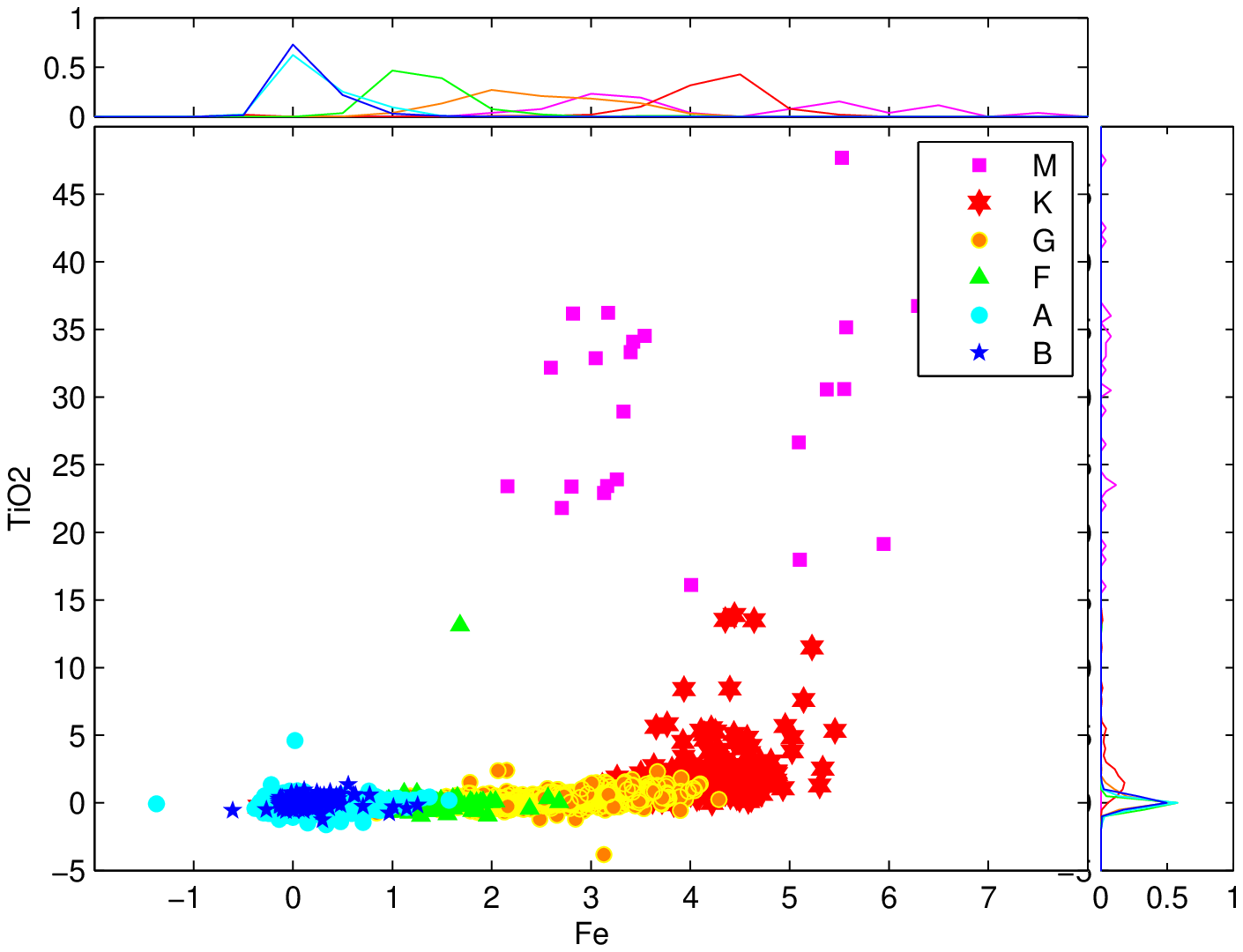}
 \end{minipage}
 \caption{The distribution of the SVM derived MK classes of the test data in the space of line indices. The top-left, top-right, middle-left, middle-right, bottom-left, and bottom-right panels show the distributions in H$_\gamma$ vs. Fe, H$_\gamma$ vs. G band, H$_\gamma$ vs. Mg, Mg vs. Fe, Fe vs. G band, and Fe vs. TiO2 planes.  The marginalized distributions of one single line index for different spectral types are shown at the right and top edges of each panel. The colors and symbols codes the OB (blue pentagons), A (cyan large circles), F (green triangles), G (orange small circles), K (red hexagons), and M (Magenta rectangles) types.}\label{fig:MKSVM}
  \end{figure}

Fig~\ref{fig:MKtrue} shows the stellar loci composed of the $\sim1500$ test dataset with color coded SIMBAD class labels in the space of line indices H$_\gamma$, Fe, Mg, G band, and TiO2. Because we use the SIMBAD MK classes as the training dataset, it implies that we assume the SIMBAD MK classes as the ``standard'' classes to be compared with. Fig~\ref{fig:MKSVM} shows the similar stellar loci with the exactly same test dataset as in Fig~\ref{fig:MKtrue}, but the colors code the SVM derived MK classes. 

Comparison between Fig~\ref{fig:MKtrue} and~\ref{fig:MKSVM} can give the qualitative impression of the performance of the SVM classification. It is obviously seen that some OB type stars (blue pentagons) located in the bottom-right corner in H$_\gamma$ vs. Fe and H$_\gamma$ vs. G4300 planes, which are shown in the two top panels in Fig~\ref{fig:MKtrue}, are mistakenly classified as A type stars (cyan circles) by the SVM method, as shown in the corresponding panels in Fig~\ref{fig:MKSVM}. Moreover, although the SVM classification works quite well for stars from M to F type, it can still see the relative harder and artificial-like boundaries among F, G, and K type stars in Fig~\ref{fig:MKSVM}.

A quantitative assessment of the performance of the SVM classification is based on the so called confusion matrix shown in Tables~\ref{tab:confusmatrix2}, in which the columns stand for the ``true" class labels and the rows stand for the SVM derived class labels. The intersections give the percentage of the stars which belong to the class in column but are assigned to the class in row by the SVM. The diagonal items show the completeness of the classification, i.e., the percentage of the stars in class X being correctly classified as the same class. The last column in Table~\ref{tab:confusmatrix2} gives the contamination, which is the percentage of the stars in the derived class X being contaminated by other classes.

Table~\ref{tab:confusmatrix2} shows that A and G type stars have the highest completeness larger than 90\%. It means that more than 90\% A or G type stars are correctly classified by the SVM algorithm. The completenesses of F and M type stars are about 72\% and 68\%, respectively, which are still acceptable. However, the completeness for OB, K, and M stars are only about 52\%, implying that almost half of these two types of stars are mis-classified in the SVM classifier. Indeed, about 44\% ``true" OB type stars are mis-classified as A type. And similar percent of ``true" K type stars are mis-classified as G type. This is probably because the spectral features of the late B (early K) type stars are very similar as those of the early A (late G) type stars and thus they are very difficult to be disentangled in SVM. It may also because that the adopted ``true" classes from SIMBAD database are compiled from various literatures and classified by eyes, and hence, not well calibrated with each other. Therefore, the large dispersions in the manually assigned MK classes may affect the performance of the SVM classification. 

\section{Discussions}\label{sect:disc}
\subsection{The discrepancy between MILES and LAMOST spectra}
\begin{figure}
 \begin{minipage}{15cm}
 \centering
 \includegraphics[scale=0.55]{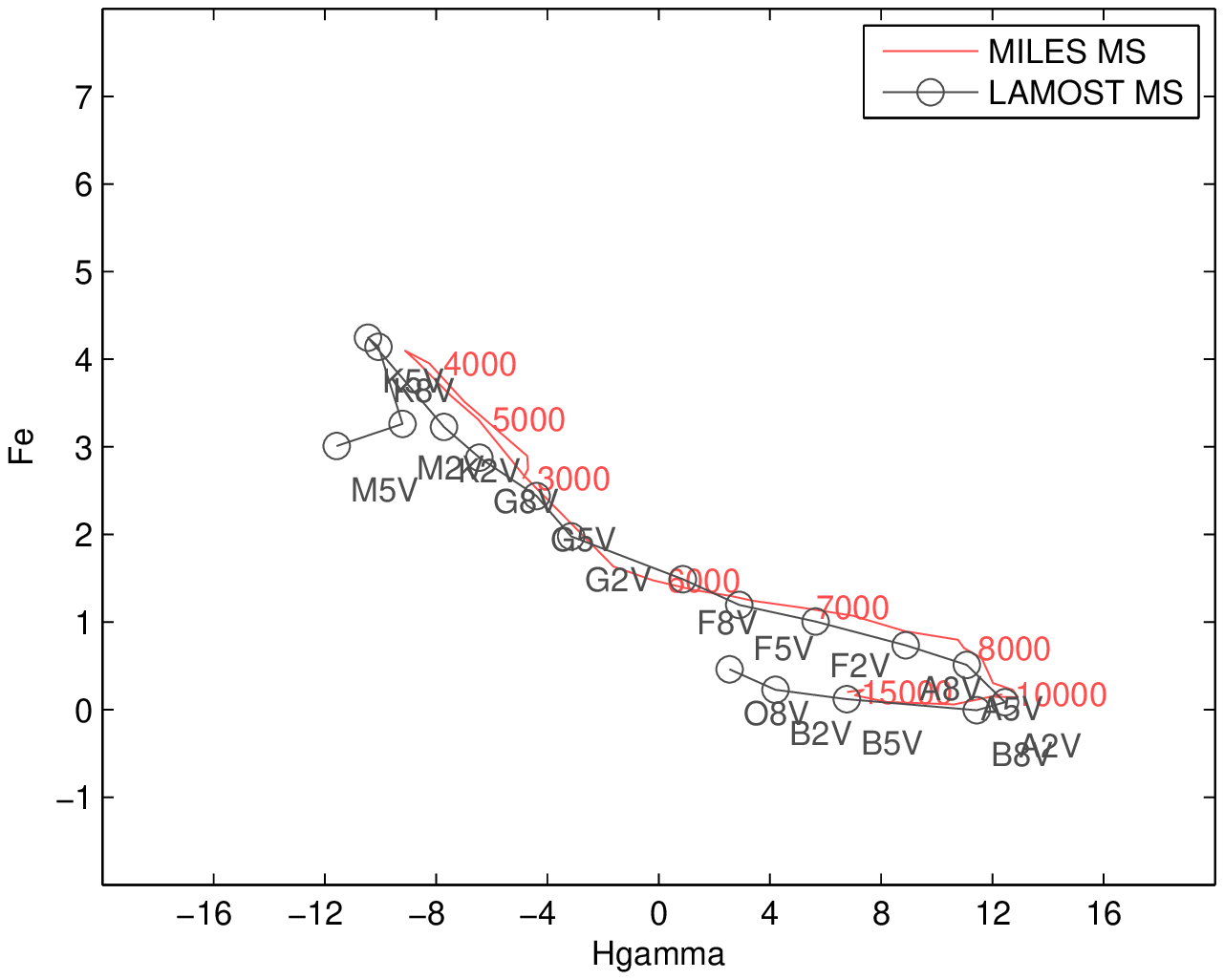}
 \includegraphics[scale=0.55]{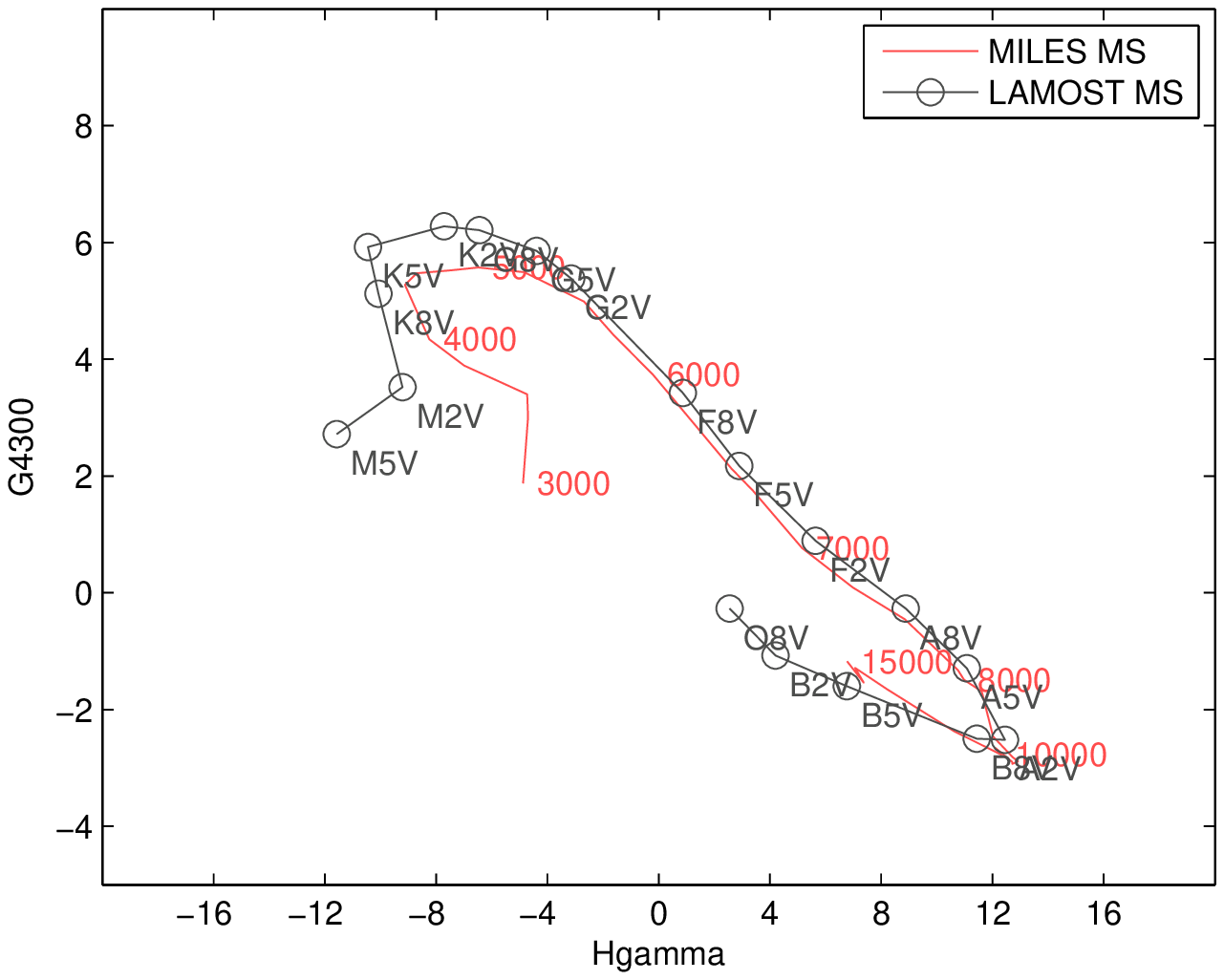}
 \includegraphics[scale=0.55]{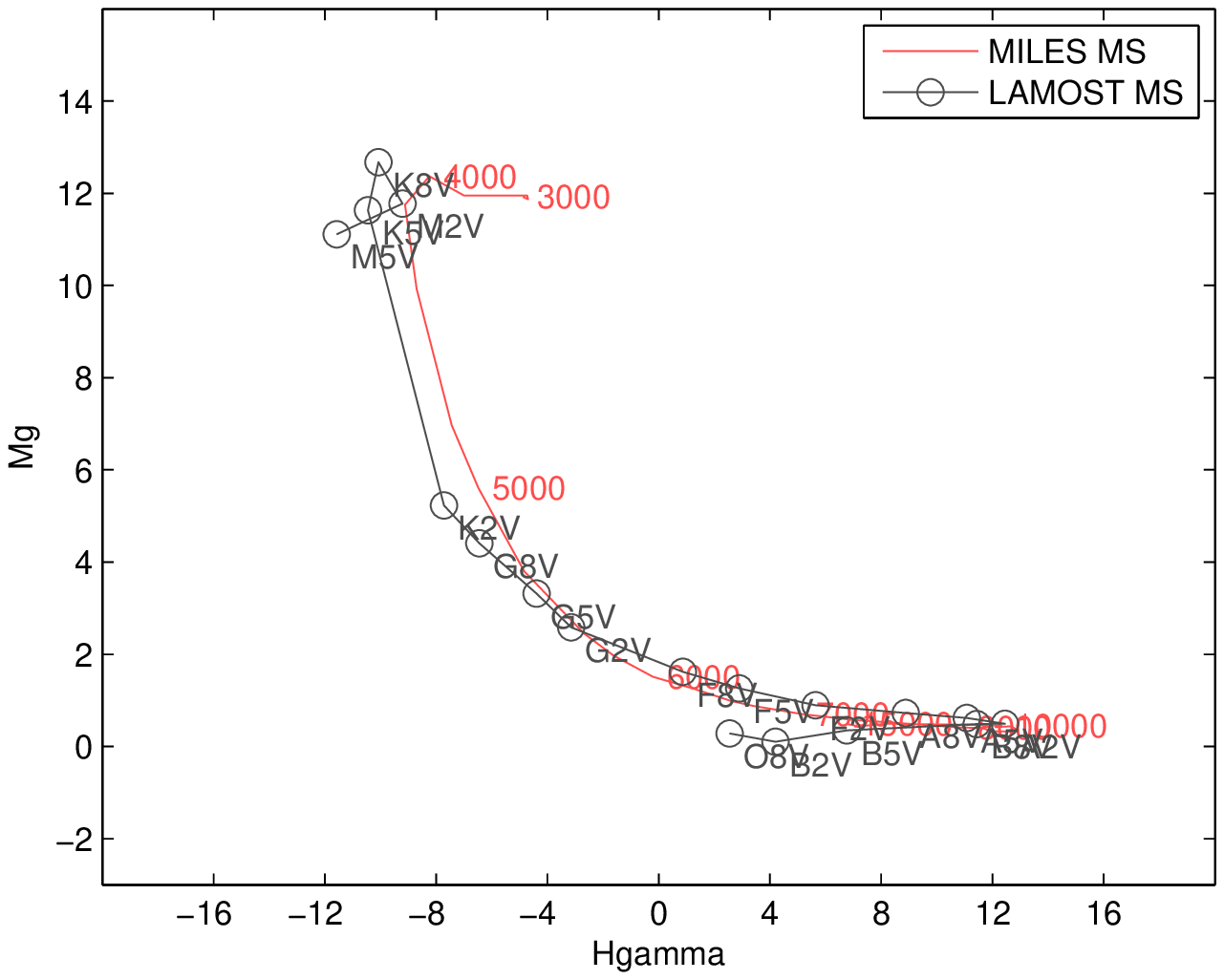}
\includegraphics[scale=0.55]{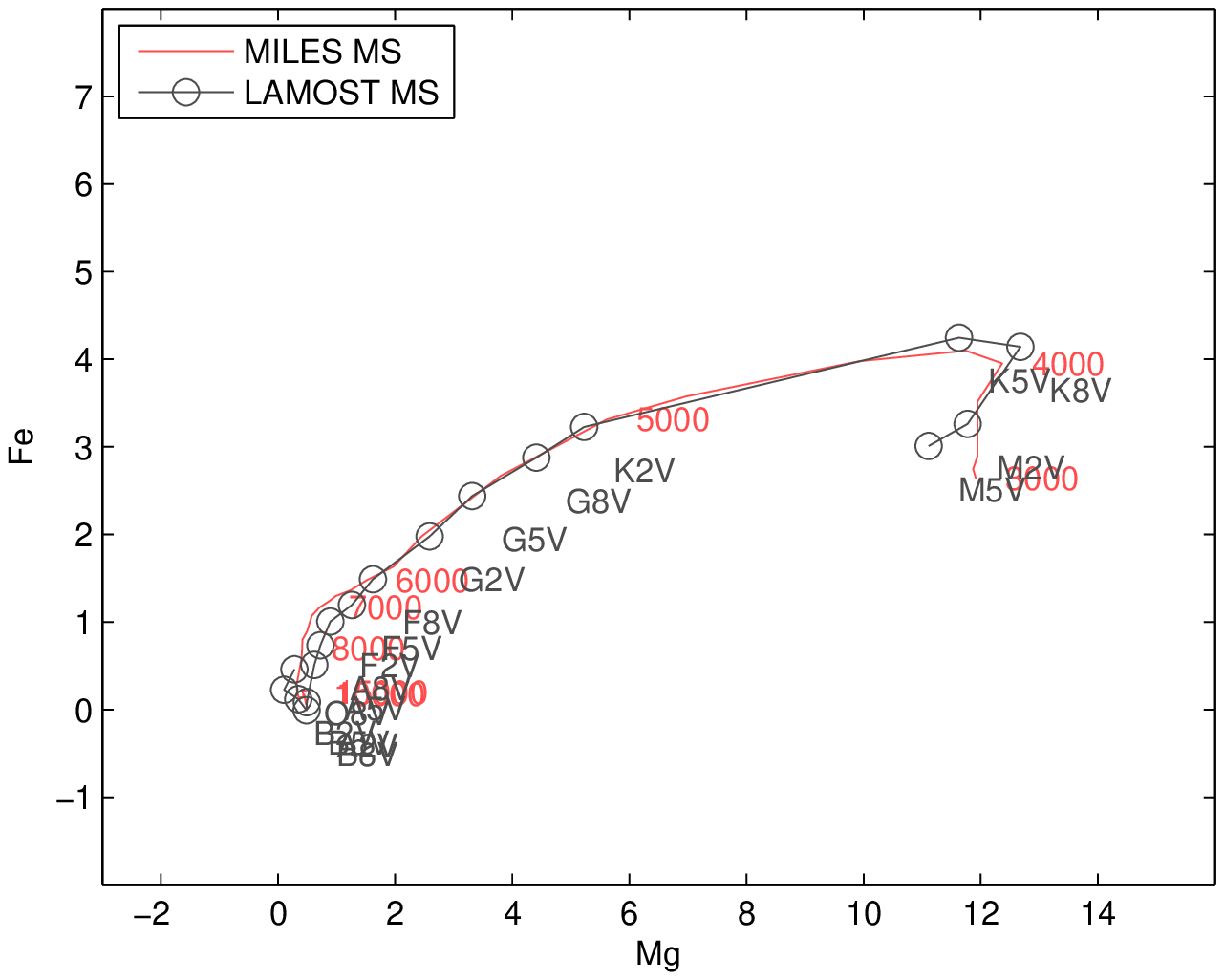}
 \includegraphics[scale=0.55]{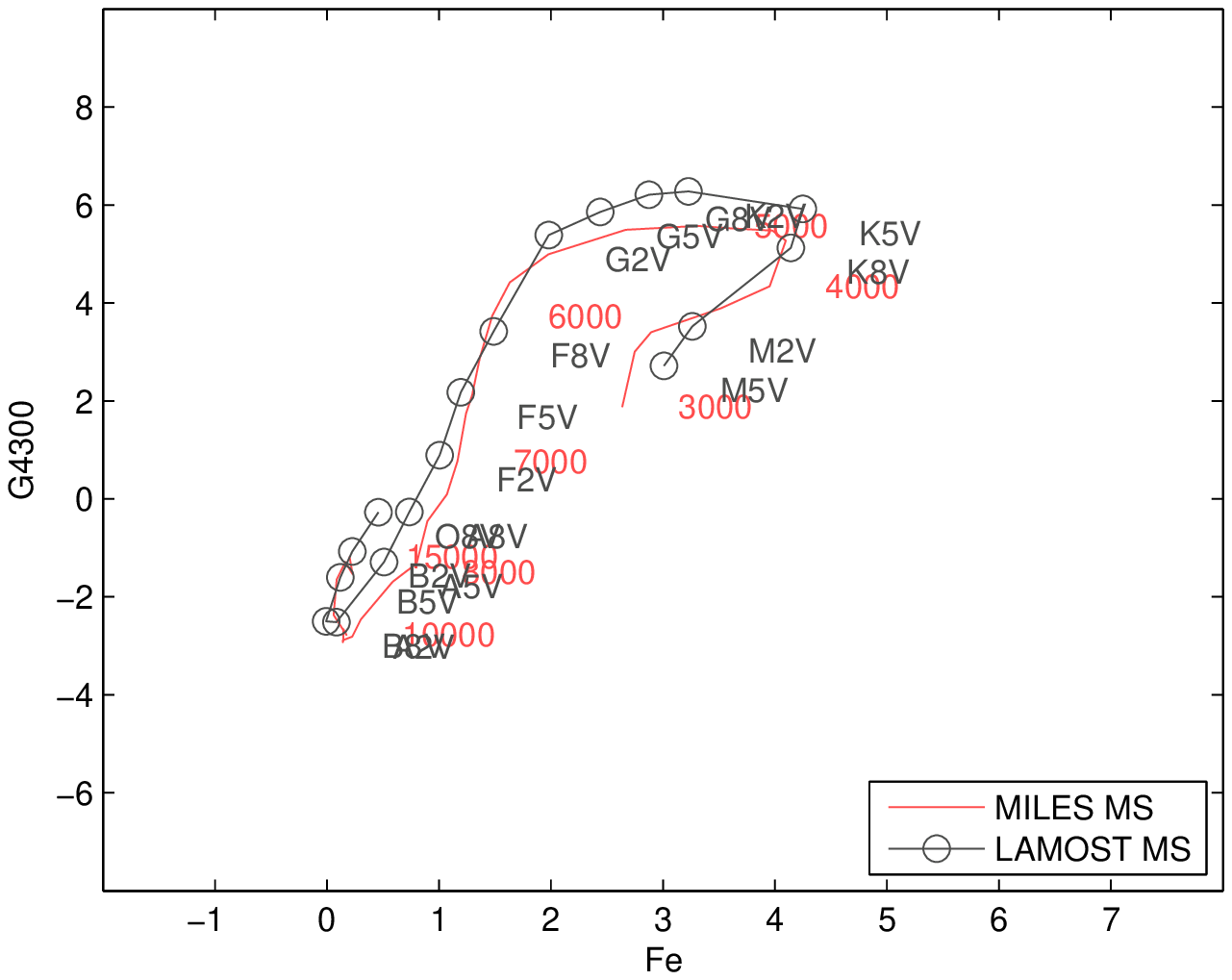}
\includegraphics[scale=0.55]{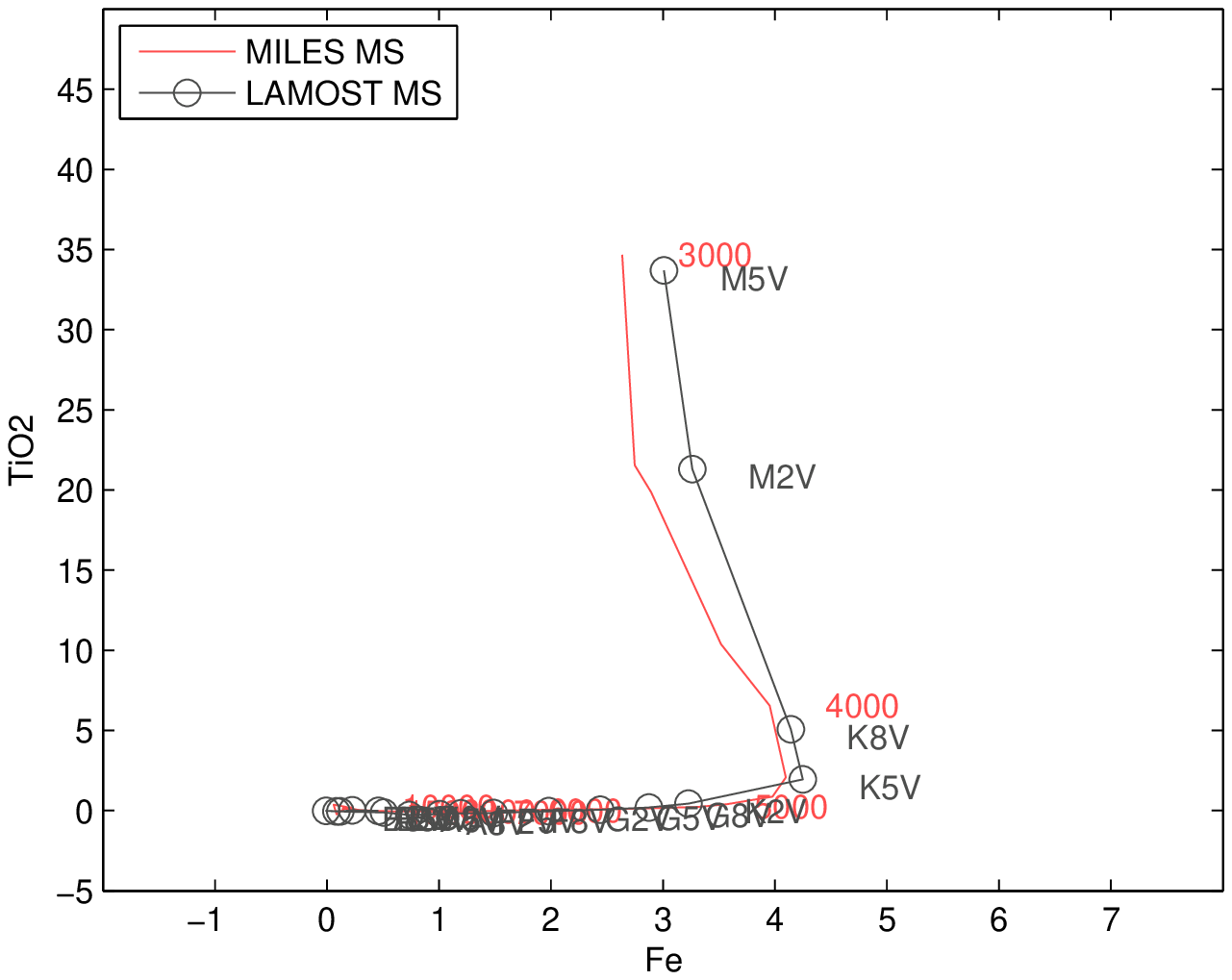}
 \end{minipage}
 \caption{The stellar loci for main-sequence (luminosity type V) stars calculated from the median location of each subtype in the space of line indices. The top-left, top-right, middle-left, middle-right, bottom-left, and bottom-right panels show the loci in H$_\gamma$ vs. Fe, H$_\gamma$ vs. G band, H$_\gamma$ vs. Mg, Mg vs. Fe, Fe vs. G band, and Fe vs. TiO2 planes. The dark lines with circles indicate the stellar loci of main-sequence stars from LAMOST spectra, while the red lines marked with the effective temperatures show the main-sequence locus of the MILES library.}\label{fig:stellarlocusMS}
\end{figure}

\begin{figure}
 \begin{minipage}{15cm}
 \centering
 \includegraphics[scale=0.55]{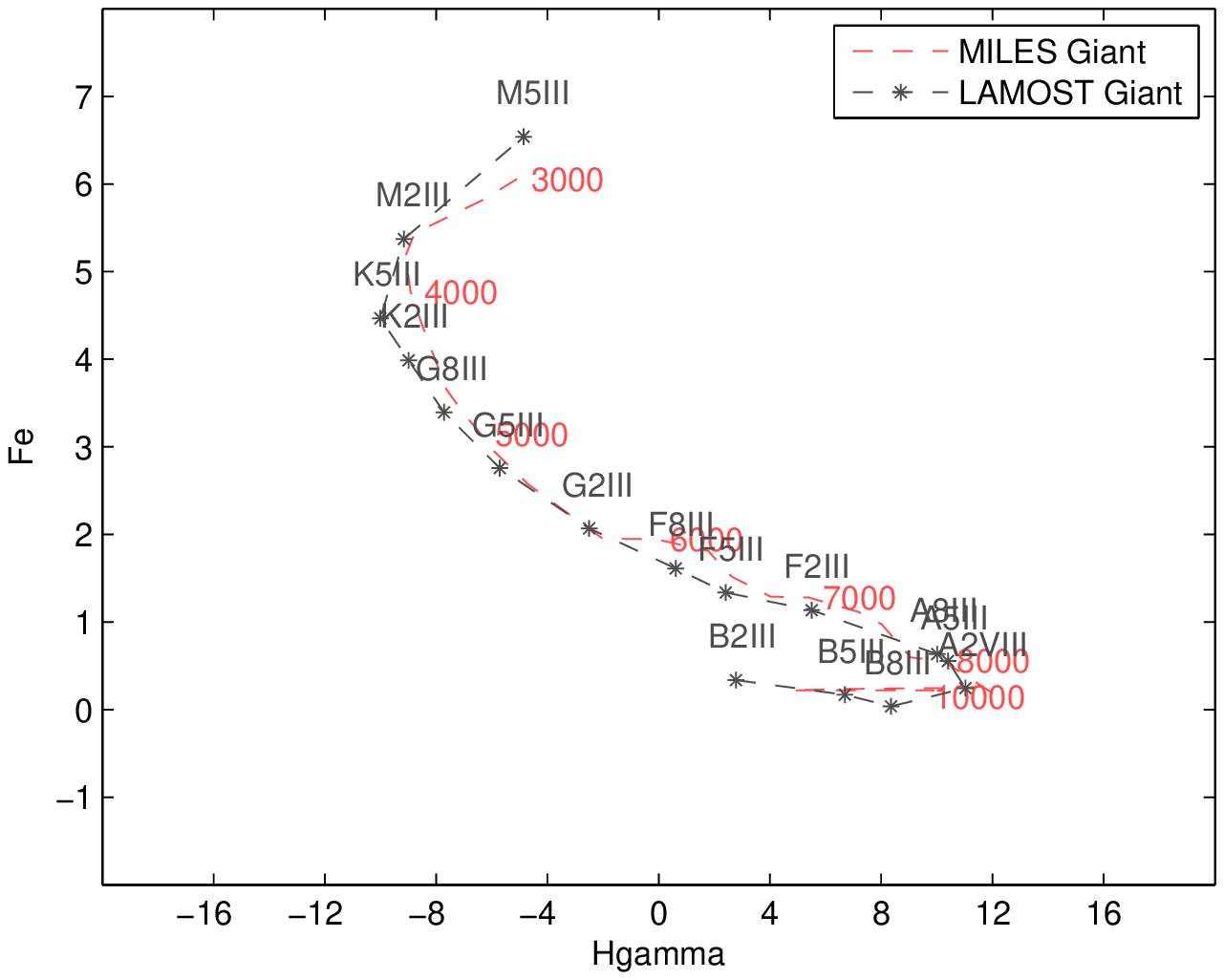}
 \includegraphics[scale=0.55]{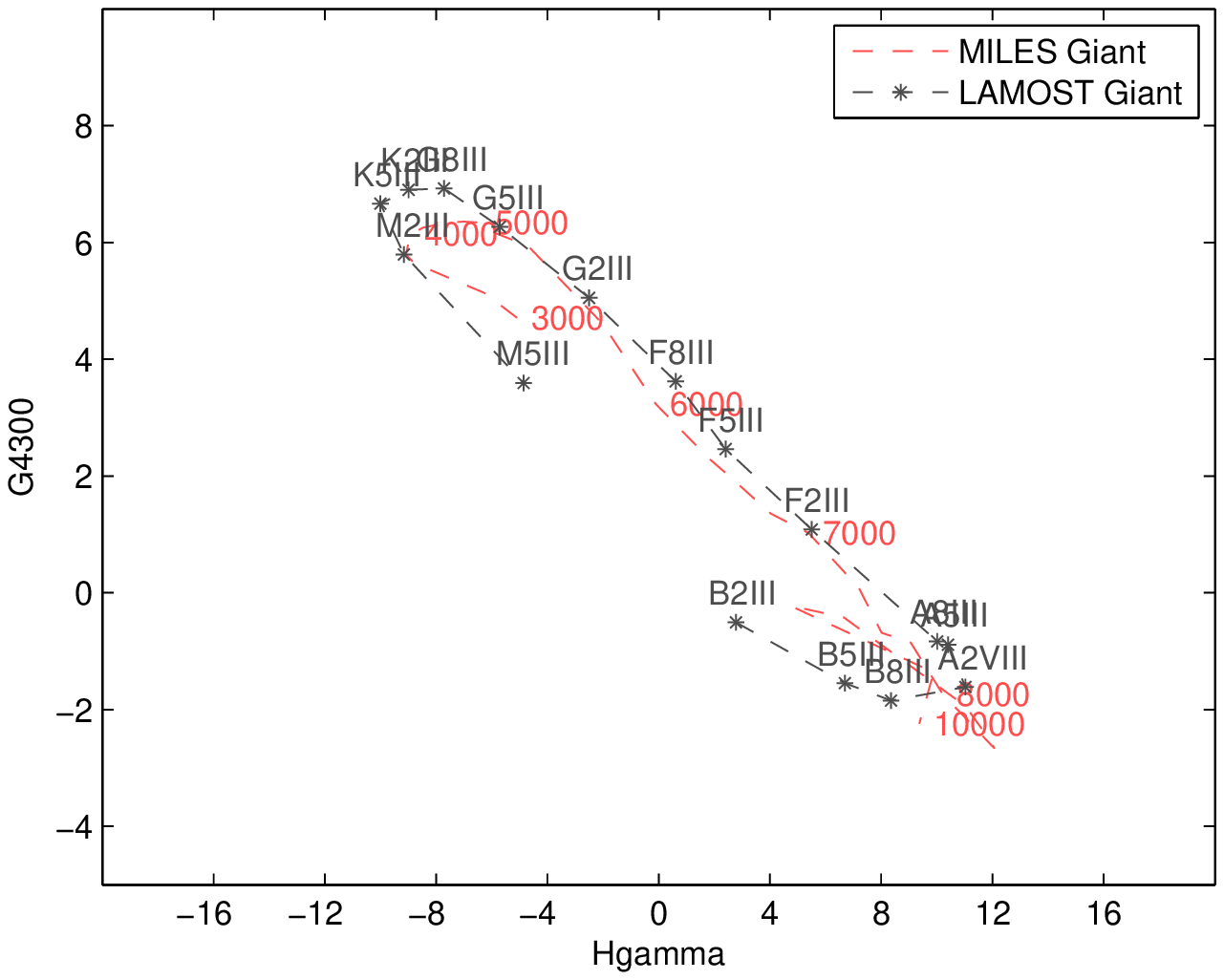}
 \includegraphics[scale=0.55]{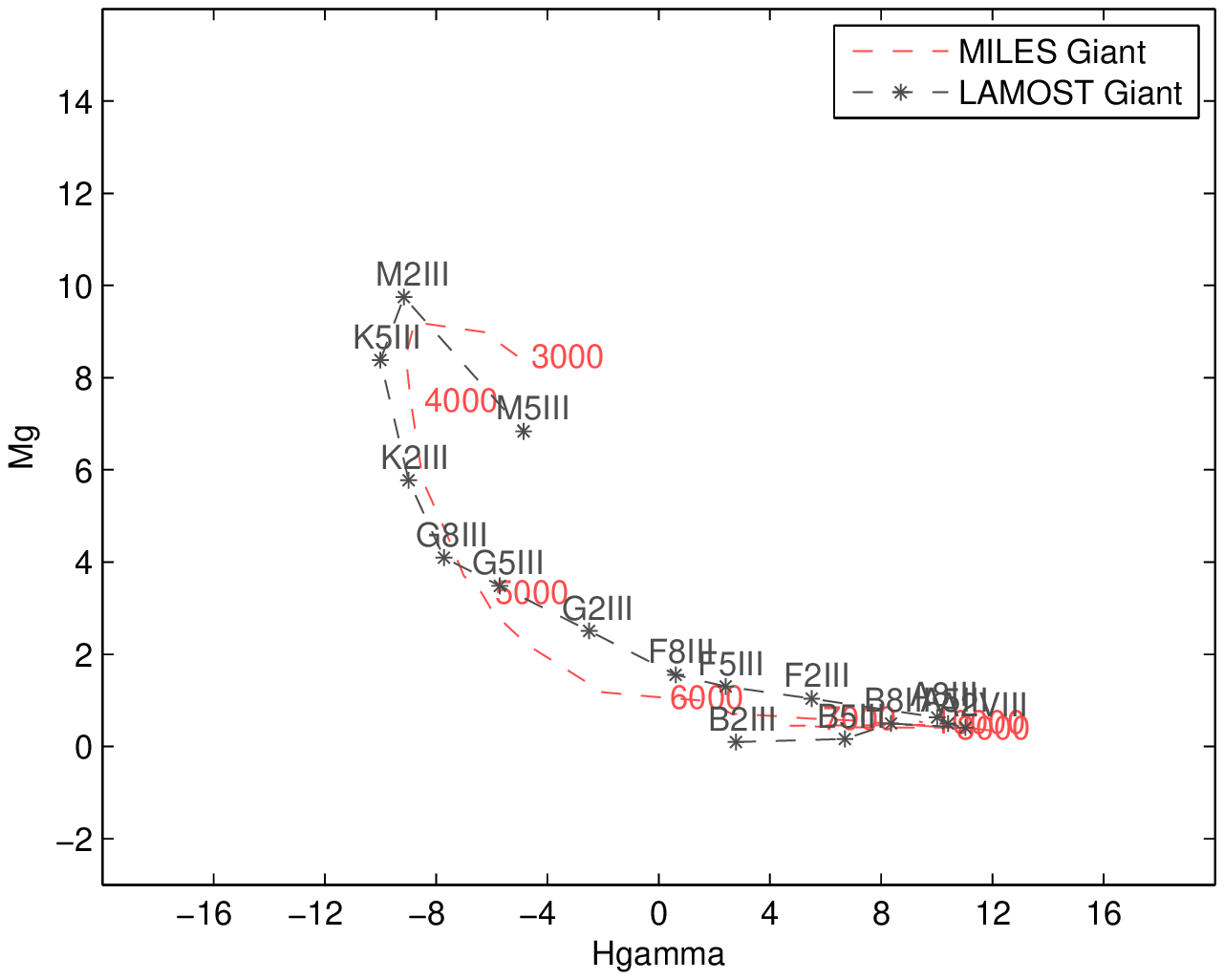}
\includegraphics[scale=0.55]{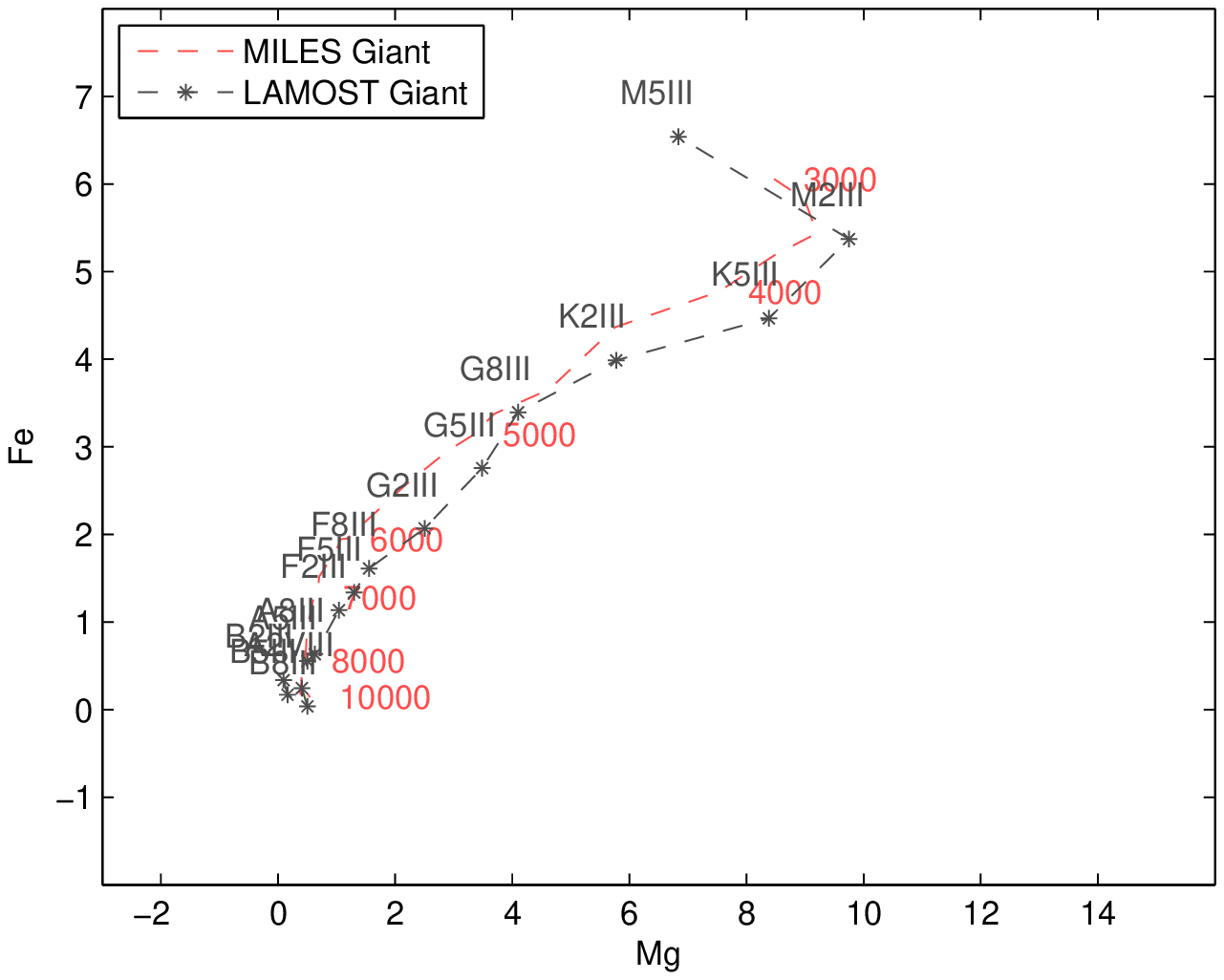}
 \includegraphics[scale=0.55]{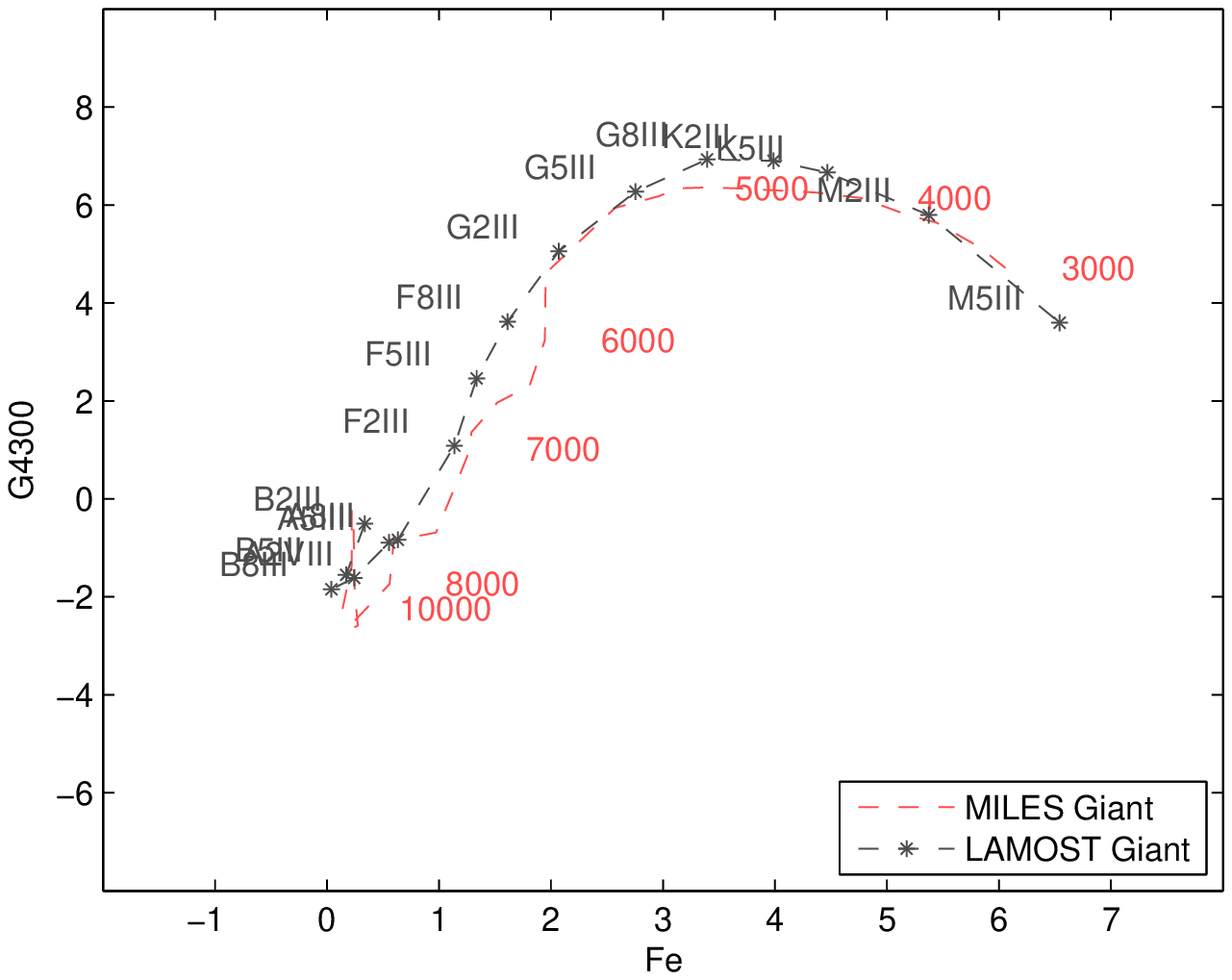}
\includegraphics[scale=0.55]{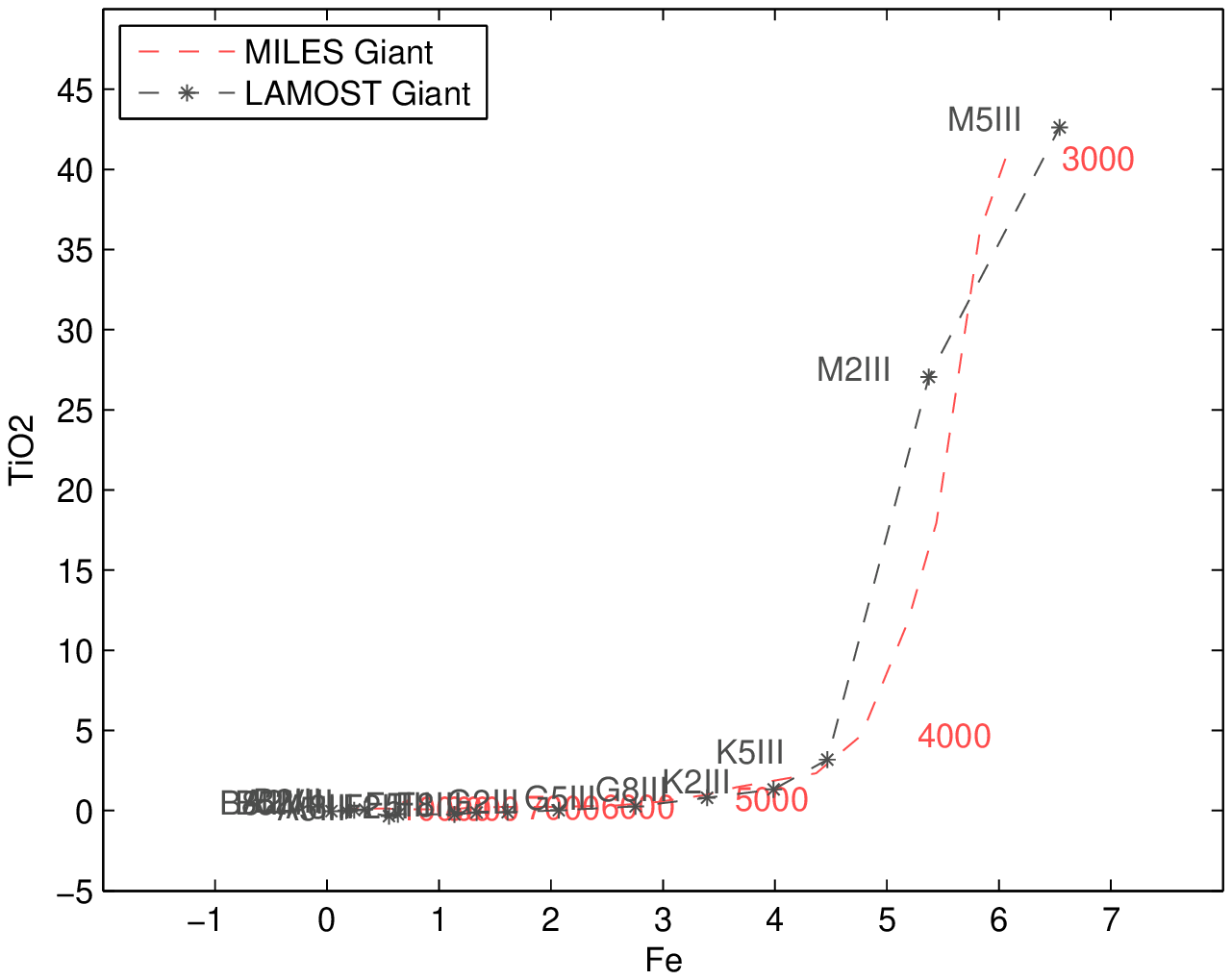}
 \end{minipage}
 \caption{The stellar loci for giant (luminosity type IV/III) stars calculated from the median location of each subtype in the space of line indices. The top-left, top-right, middle-left, middle-right, bottom-left, and bottom-right panels show the loci in H$_\gamma$ vs. Fe, H$_\gamma$ vs. G band, H$_\gamma$ vs. Mg, Mg vs. Fe, Fe vs. G band, and Fe vs. TiO2 planes. The dashed lines with asterisks indicate the stellar loci of giant stars from LAMOST spectra, while the red dashed lines marked with the effective temperatures show the giant locus of the MILES library.}\label{fig:stellarlocusGNT}
 \end{figure}

In order to provide an external comparison of the stellar locus in the space of the line indices, we calculate the same line indices for the MILES samples (S\'anchez-Bl\'azquez et al.~\cite{miles}), which contains 985 bright stellar spectra with wide extensions in stellar parameters. We overlap the stellar loci of MILES data with red lines in
Figs~\ref{fig:stellarlocusMS} and~\ref{fig:stellarlocusGNT} for main-sequence and giant stars, respectively. To be convenient, we also mark the averaged effective temperatures along the stellar loci as the reference. They show that the stellar loci of LAMOST and MILES are not completely overlapped with each other, especially for late type dwarf stars and all giant stars. Although, according to Tables~\ref{tab:MSlocus} and~\ref{tab:giantlocus}, these differences are mostly within uncertainty of 1- or 2-$\sigma$, the overall shifts of the loci of MILES in most panels of Figs~\ref{fig:stellarlocusMS} and~\ref{fig:stellarlocusGNT} are likely systematic. {Looking back to the bottom-right panel of Fig~\ref{fig:stellarlocusMSGNT}, it is seen that the SIMBAD stellar locus for M dwarf stars show similar systematic bias from the full sample of the LAMOST data (the contours). Therefore, it is likely that the M dwarf stars in the SIMBAD database may be a biased sample and cannot represent for the majority of the LAMOST M dwarf samples.}
This also gives an alert that the line indices stellar loci derived from one survey should not be directly extended to other survey. Calibrations in the line indices and in the sample selection function are necessary before the extension. 

\subsection{How to make the decision, the MK class or the line indices-based stellar locus?}
In the previous sections we show two kinds of classifications. The line indices stellar locus orders the different types of stars as a simple sequence, along which the effective temperature monotonically changes from coolest to hottest. No hard boundary has to be set in the stellar locus to artificially separate the stars into discrete classes. The users who want to select specific stars for their statistical studies on the Milky Way can simply cut the data from any segment of the stellar locus. 

On the other hand, the SVM based classification assigns discrete MK type labels to stars based on the prior knowledge---the SIMBAD MK class labels. The compiled MK classes in SIMBAD database are from lots of literatures, most of them are done by comparing the spectra with the small sample of the standard stars by eyes. This may raise significant inconsistency between the literatures. Calibrations among different literature seems very difficult, since the MK classes are not continuous but discrete. 

The realistic issue for large spectroscopic surveys, such as the LAMOST survey, is that millions of the stars are observed and it is impossible to inspect each spectrum by eyes. As shown in the exercise of SVM classification in section~\ref{sect:svm}, the state-of-the-art machine learning techniques may not very helpful because they need to be trained by the prior knowledge which should be accurate and self-consistent.

Based on this analysis, we therefore suggest the LAMOST users to employ the line indices stellar locus, rather than directly use the derived MK classes from the catalog, to select the proper types of stars to meet their specific request. If the users want to compare their sample with literatures, which may use MK classes, they can quantitatively calculate the percentage of completeness and contaminations via the comparison of the stellar loci with SIMBAD and the SVM MK classes.

\section{Conclusions}\label{sect:conc}
In this paper, we revisit the fundamental issue of the stellar classification using 3,000 high signal-to-noise ratio LAMOST spectra with known MK classes obtained from the cross-identification of SIMBAD database. Although the MK classes have been widely used for more than 70 years and become a standard, it seems not easy to adapt the large amount data from precent-day spectroscopic surveys. The MK classes are constructed based on a very small sample of standard stars, which are mostly very bright and located in the local volume nearby the Sun. New spectroscopic surveys, e.g.,  SDSS and LAMOST, can detect the deep sky as far as 100\,kpc and hence contains millions of stars from very different populations with the solar neighborhood. The current standard star library then becomes incomplete compared with a few orders of magnitude larger survey data. Another issue is that almost all stars with known MK classes are classified by eyes. It is unfortunately impossible in the era of large data. The third issue is that the MK classes are discrete, which make it difficult to be calibrated.

We map the MK classes into the space of line indices and find that the stellar loci in the lines indices can well describe the MK classes. Moreover, it is naturally along the change of the effective temperature. For the late type stars, the different luminosity types can also be disentangled in the stellar loci. 

We then investigate the performance of an automatic MK classification based on the SVM technique. We find that although A, F, G, and M types of stars can be well classified, almost half of the B or K type stars are mis-classified. 

We therefore suggest that the classification of the stars should be based on the continuous stellar loci in line indices. The advantages of the stellar loci are that 1) they are continuous and one can cut a group of data at any point on the loci; 2) the stellar loci is consistent with the effective temperature; and 3) after selecting a group of stars from the stellar loci, one can easily estimate the completeness and contamination of the sample in terms of MK classes.

\begin{acknowledgements}
This work is supported by the Strategic Priority Research Program ``The Emergence of Cosmological Structures" of the Chinese Academy of Sciences, Grant No. XDB09000000 and the National Key Basic Research Program of China 2014CB845700. CL acknowledges the National Science Foundation of China (NSFC) under grants 11373032, 11333003 and U1231119. Guoshoujing Telescope (the Large Sky Area Multi-Object Fiber Spectroscopic Telescope LAMOST) is a National Major Scientific Project built by the Chinese Academy of Sciences. Funding for the project has been provided by the National Development and Reform Commission. LAMOST is operated and managed by the National Astronomical Observatories, Chinese Academy of Sciences.
\end{acknowledgements}


\clearpage
\begin{table}
\caption{The number of the MK classes for the LAMOST-SIMBAD sample}\label{tab:sample}
\begin{center}
\begin{tabular}{l|c|c|c|c||l|c|c|c|c}
\hline\hline
Type & Total & V & IV/III & II/I & Type & Total & V & IV/III & II/I\\
\hline
O5 & 1 & 1 & 0 & 0 & F7 & 15 & 12 & 2 & 1\\
O7 & 2 & 2 & 0 & 0 & F8 & 55 & 38 & 15 & 2\\
O8 & 1 & 1 & 0 & 0 & F9 & 22 & 11 & 11 & 0\\
O9 & 4 & 3 & 1 & 0 & G0 & 435 & 398 & 36 & 1\\
B0 & 14 & 9 & 5 & 0 & G1 & 21 & 19 & 2 & 0\\
B1 & 15 & 8 & 7 & 0 & G2 & 57 & 33 & 24 & 0\\
B2 & 19 & 11 & 8 & 0 & G3 & 16 & 9 & 7 & 0\\
B3 & 9 & 7 & 0 & 2 & G4 & 21 & 17 & 2 & 2\\
B4 & 9 & 7 & 2 & 0 & G5 & 280 & 218 & 62 & 0\\
B5 & 34 & 18 & 15 & 1 & G6 & 23 & 11 & 12 & 0\\
B6 & 3 & 2 & 0 & 1 & G7 & 17 & 7 & 10 & 0\\
B7 & 23 & 13 & 10 & 0 & G8 & 224 & 114 & 109 & 1\\
B8 & 75 & 65 & 10 & 0 & G9 & 27 & 8 & 19 & 0\\
B9 & 175 & 157 & 18 & 0 & K0 & 183 & 89 & 74 & 3\\
A0 & 420 & 386 & 30 & 4 & K1 & 56 & 13 & 31 & 0\\
A1 & 67 & 63 & 4 & 0 & K2 & 83 & 38 & 33 & 0\\
A2 & 186 & 175 & 10 & 1 & K3 & 25 & 15 & 7 & 0\\
A3 & 61 & 60 & 1 & 0 & K4 & 25 & 17 & 8 & 0\\
A4 & 11 & 10 & 0 & 1 & K5 & 21 & 14 & 5 & 1\\
A5 & 43 & 39 & 2 & 2 & K6 & 9 & 9 & 0 & 0\\
A6 & 3 & 2 & 1 & 0 & K7 & 11 & 11 & 0 & 0\\
A7 & 27 & 21 & 6 & 0 & K8 & 5 & 5 & 0 & 0\\
A8 & 12 & 10 & 0 & 2 & K9 & 2 & 2 & 0 & 0\\
A9 & 1 & 0 & 1 & 0 & M0 & 21 & 12 & 9 & 0\\
F0 & 53 & 34 & 16 & 3 & M1 & 6 & 3 & 3 & 0\\
F1 & 4 & 1 & 3 & 0 & M2 & 13 & 9 & 4 & 0\\
F2 & 36 & 27 & 8 & 1 & M3 & 12 & 7 & 5 & 0\\
F3 & 14 & 10 & 4 & 0 & M4 & 10 & 9 & 1 & 0\\
F4 & 7 & 5 & 1 & 1 & M5 & 4 & 2 & 2 & 0\\
F5 & 67 & 54 & 12 & 1 & M7 & 2 & 0 & 1 & 1\\
F6 & 36 & 21 & 14 & 1 & M8 & 1 & 1 & 0 & 0\\
\hline\hline
\end{tabular}
\end{center}
\end{table}

\begin{table*}
\begin{center}
\caption{Line indices definition}\label{tab:lineindices}
\begin{tabular}{c|c|c}
\hline\hline
Name & Index Bandpass ($\rm\AA$) & Pseudocontinua ($\rm\AA$)\\
\hline
CaII K$^{\rm a}$ & 3927.7-3939.7 & 3903-3923 4000-4020\\
H$_{\delta}$$^{\rm b}$ & 4083.50-4122.25 & 4041.60-4079.75 4128.50-4161.00\\
CN$^{\rm c}$ & 4143.375-4178.375 & 4081.375-4118.875 4245.375-4285.375\\
Ca4227$^{\rm c}$ & 4223.500-4236.000&4212.250-4221.000 4242.250-4252.250\\
G4300$^{\rm c}$ & 4282.625-4317.625&4267.625-4283.875 4320.125-4336.375\\
H$_{\gamma}$$^{\rm b}$ &4319.75-4363.50&4283.50-4319.75 4367.25-4419.75\\
Fe4383$^{\rm c}$&4370.375-4421.625&4360.375-4371.625 4444.125-4456.625\\
He4388&4381-4399&4365-4380 4398-4408\\
Ca4455$^{\rm c}$&4453.375-4475.875&4447.125-4455.875 4478.375-4493.375\\
He4471&4462-4475&4450-4463 4485-4495\\
Fe4531$^{\rm c}$&4515.500-4560.500&4505.500-4515.500 4561.750-4580.500\\
He4542&4536-4548&4526-4536 4548-4558\\
Fe4668$^{\rm c}$&4635.250-4721.500&4612.750-4631.500 4744.000-4757.750\\
H$_{\beta}$$^{\rm b}$ &4847.875-4876.625&4827.875-4847.875 4876.625-4891.625\\
Fe5015$^{\rm c}$&4977.750-5054.000&4946.500-4977.750 5054.000-5065.250\\
Mg$_1$$^{\rm c}$&5069.125-5134.125&4895.125-4957.625 5301.125-5366.125\\
Mg$_2$$^{\rm c}$&5154.125-5196.625&4895.125-4957.625 5301.125-5366.125\\
Mg$_b$$^{\rm c}$&5160.125-5192.625&5142.625-5161.375 5191.375-5206.375\\
Fe5270$^{\rm c}$&5245.650-5285.650&5233.150-5248.150 5285.650-5318.150\\
Fe5335$^{\rm c}$&5312.125-5352.125&5304.625-5315.875 5353.375-5363.375\\
Fe5406$^{\rm c}$&5387.500-5415.000&5376.250-5387.500 5415.000-5425.000\\
Fe5709$^{\rm c}$&5698.375-5722.125&5674.625-5698.375 5724.625-5738.375\\
Fe5782$^{\rm c}$&5778.375-5798.375&5767.125-5777.125 5799.625-5813.375\\
NaD$^{\rm c}$&5878.625-5911.125&5862.375-5877.375 5923.875-5949.875\\
TiO$_1$$^{\rm c}$&5938.375-5995.875&5818.375-5850.875 6040.375-6105.375\\
TiO2$^{\rm c}$&6191.375-6273.875&6068.375-6143.375 6374.375-6416.875\\
H$_{\alpha}$$^{\rm d}$&6548.00-6578.00&6420.00-6455.00 6600.00-6640.00\\
\hline\hline
\end{tabular}
\end{center}
$^{\rm a}$\ Beers et al.~\cite{beers99}\\
$^{\rm b}$\ Worthey \& Ottaviani~\cite{worthey97}\\
$^{\rm c}$\ Worthey et al.~\cite{worthey94}\\
$^{\rm d}$\ Cohen, Blakeslee \& Ryzhov~\cite{cohen98}\\
\end{table*}

\begin{table}
\caption{The median locus in the space of equivalent width of the spectral lines for main-sequence stars.}\label{tab:MSlocus}
\centering
\begin{tabular}{l|c|c|c|c|c|c}
\hline\hline
 Type & $EW_{G4300}$ & $EW_{H\gamma}$ & $EW_{Mg}$ & $EW_{Fe}$ & $EW_{TiO2}$ & Number\\
& $\rm\AA$ & $\rm\AA$ & $\rm\AA$ & $\rm\AA$ & $\rm\AA$ & of stars\\
 \hline
O6-9 & -0.27$\pm$0.23 & 2.55$\pm$0.23 & 0.46$\pm$0.12 & 0.28$\pm$0.43 & -0.00$\pm$0.28 & 6\\
B0-3 & -1.07$\pm$0.45 & 4.20$\pm$1.72 & 0.22$\pm$0.12 & 0.10$\pm$0.35 & 0.03$\pm$0.49 & 35\\
B3-6 & -1.60$\pm$0.59 & 6.76$\pm$1.62 & 0.12$\pm$0.11 & 0.35$\pm$0.28 & -0.04$\pm$0.33 & 34\\
B6-9 & -2.50$\pm$1.04 & 11.43$\pm$2.68 & -0.01$\pm$0.28 & 0.49$\pm$1.96 & -0.01$\pm$5.37 & 237\\
A0-3 & -2.52$\pm$1.32 & 12.45$\pm$2.65 & 0.08$\pm$0.58 & 0.49$\pm$0.90 & -0.05$\pm$0.43 & 684\\
A3-6 & -1.29$\pm$1.01 & 11.07$\pm$2.09 & 0.51$\pm$0.33 & 0.62$\pm$0.29 & -0.11$\pm$0.40 & 111\\
A6-9 & -0.27$\pm$0.80 & 8.88$\pm$1.86 & 0.74$\pm$0.30 & 0.72$\pm$0.30 & -0.27$\pm$83.98 & 33\\
F0-3 & 0.89$\pm$1.04 & 5.64$\pm$2.28 & 1.01$\pm$0.38 & 0.89$\pm$0.31 & -0.21$\pm$7.70 & 72\\
F3-6 & 2.18$\pm$1.20 & 2.90$\pm$2.75 & 1.19$\pm$0.36 & 1.26$\pm$0.56 & -0.17$\pm$7.61 & 90\\
F6-9 & 3.42$\pm$1.11 & 0.87$\pm$2.16 & 1.49$\pm$0.39 & 1.62$\pm$0.50 & -0.15$\pm$3.48 & 82\\
G0-3 & 5.38$\pm$1.08 & -3.15$\pm$2.29 & 1.98$\pm$0.61 & 2.59$\pm$1.93 & -0.03$\pm$0.48 & 459\\
G3-6 & 5.86$\pm$0.98 & -4.39$\pm$2.45 & 2.44$\pm$0.67 & 3.32$\pm$2.42 & 0.05$\pm$0.49 & 255\\
G6-9 & 6.21$\pm$0.72 & -6.45$\pm$1.97 & 2.88$\pm$0.76 & 4.41$\pm$1.45 & 0.19$\pm$0.46 & 140\\
K0-3 & 6.28$\pm$1.35 & -7.72$\pm$3.53 & 3.23$\pm$0.91 & 5.23$\pm$2.42 & 0.45$\pm$2.52 & 155\\
K3-6 & 5.92$\pm$0.84 & -10.45$\pm$2.78 & 4.25$\pm$0.55 & 11.63$\pm$2.49 & 1.96$\pm$11.10 & 55\\
K6-9 & 5.12$\pm$0.94 & -10.07$\pm$4.24 & 4.14$\pm$0.71 & 12.68$\pm$1.33 & 5.08$\pm$15.70 & 27\\
M0-3 & 3.52$\pm$1.31 & -9.21$\pm$3.28 & 3.26$\pm$0.81 & 11.78$\pm$0.91 & 21.30$\pm$15.85 & 31\\
M3-6 & 2.72$\pm$1.02 & -11.57$\pm$5.31 & 3.01$\pm$0.45 & 11.11$\pm$2.66 & 33.69$\pm$20.42 & 18\\
\hline\hline
\end{tabular}
\end{table}

\begin{table}
\caption{The median locus in the space of equivalent width of the spectral lines for stars with luminosity types IV or III.}\label{tab:giantlocus}
\centering
\begin{tabular}{l|c|c|c|c|c|c}
\hline\hline
 Type & $EW_{G4300}$ & $EW_{H\gamma}$ & $EW_{Mg}$ & $EW_{Fe}$ & $EW_{TiO2}$ & Number\\
& $\rm\AA$ & $\rm\AA$ & $\rm\AA$ & $\rm\AA$ & $\rm\AA$ & of stars\\
 \hline
B0-3 & -0.51$\pm$0.34 & 2.78$\pm$1.78 & 0.34$\pm$0.21 & 0.09$\pm$0.33 & 0.07$\pm$13.83 & 20\\
B3-6 & -1.55$\pm$0.59 & 6.70$\pm$1.33 & 0.17$\pm$0.21 & 0.16$\pm$0.32 & 0.01$\pm$0.34 & 17\\
B6-9 & -1.85$\pm$1.04 & 8.36$\pm$2.35 & 0.04$\pm$0.35 & 0.50$\pm$0.34 & -0.02$\pm$13.38 & 38\\
A0-3 & -1.61$\pm$0.95 & 11.03$\pm$1.97 & 0.24$\pm$0.40 & 0.41$\pm$0.36 & 0.03$\pm$0.32 & 45\\
A3-6 & -0.89$\pm$1.12 & 10.41$\pm$1.75 & 0.55$\pm$0.28 & 0.50$\pm$0.20 & -0.35$\pm$0.08 & 4\\
A6-9 & -0.83$\pm$1.16 & 10.02$\pm$2.17 & 0.63$\pm$0.32 & 0.63$\pm$0.23 & -0.22$\pm$0.18 & 8\\
F0-3 & 1.09$\pm$1.10 & 5.50$\pm$2.58 & 1.14$\pm$0.21 & 1.04$\pm$0.36 & -0.24$\pm$14.86 & 31\\
F3-6 & 2.46$\pm$0.91 & 2.41$\pm$1.80 & 1.34$\pm$0.31 & 1.30$\pm$0.45 & -0.11$\pm$11.07 & 31\\
F6-9 & 3.62$\pm$1.03 & 0.61$\pm$2.01 & 1.61$\pm$0.33 & 1.56$\pm$0.50 & -0.10$\pm$2.06 & 42\\
G0-3 & 5.05$\pm$1.28 & -2.50$\pm$2.91 & 2.07$\pm$0.81 & 2.50$\pm$1.49 & 0.03$\pm$9.27 & 69\\
G3-6 & 6.27$\pm$1.17 & -5.71$\pm$2.78 & 2.76$\pm$0.78 & 3.49$\pm$1.42 & 0.27$\pm$0.61 & 83\\
G6-9 & 6.93$\pm$0.91 & -7.72$\pm$2.15 & 3.40$\pm$2.55 & 4.10$\pm$3.74 & 0.80$\pm$10.62 & 150\\
K0-3 & 6.90$\pm$0.88 & -8.99$\pm$3.16 & 3.99$\pm$0.77 & 5.77$\pm$2.14 & 1.32$\pm$4.82 & 189\\
K3-6 & 6.67$\pm$0.93 & -10.01$\pm$3.38 & 4.47$\pm$0.80 & 8.38$\pm$2.20 & 3.18$\pm$3.80 & 24\\
M0-3 & 5.79$\pm$1.42 & -9.16$\pm$3.06 & 5.37$\pm$0.87 & 9.75$\pm$1.87 & 27.04$\pm$13.31 & 21\\
M3-6 & 3.59$\pm$0.99 & -4.85$\pm$9.35 & 6.54$\pm$0.67 & 6.84$\pm$1.67 & 42.62$\pm$6.16 & 8\\
 \hline\hline
 \end{tabular}
 \end{table}
 
 
 \begin{table}
 \caption{The confusion matrix in percentage of the SVM-based MK classification}\label{tab:confusmatrix2}
 \begin{center}
 \begin{tabular}{l|l|c|c|c|c|c|c|c}
    \hline\hline
      &  &\multicolumn{6}{|c|}{SIMBAD}&\\
    \hline
 &  & OB & A & F & G & K& M &contamination\\
\hline
 \multirow{5}{*}{SVM}& OB   & 52.60\% & 6.97\% & 1.94\% & 0.00\% & 0.00\% & 0.00\% & 24.06\%\\
 & A  & 44.79\% & 90.38\% & 8.39\% & 0.53\% & 1.43\% & 0.00\% & 21.83\%\\
 & F   & 1.56\% & 1.68\% & 72.26\% & 3.57\% & 0.95\% & 0.00\% & 22.22\%\\
 & G  & 0.52\% & 0.48\% & 17.42\% & 90.91\% & 43.81\% & 2.86\% & 19.43\%\\
 & K  & 0.52\% & 0.48\% & 0.00\% & 4.99\% & 52.86\% & 28.57\% & 26.97\%\\
 & M  & 0.00\% & 0.00\% & 0.00\% & 0.00\% & 0.95\% & 68.57\% & 7.69\%\\  
    \hline\hline
 \end{tabular}
 \end{center}
 \end{table}
 
 \clearpage

\label{lastpage}

\end{document}